\documentclass{aa}  

\usepackage{graphicx}
\usepackage{txfonts}
\usepackage{multirow}
\usepackage{colortbl}

\usepackage{hyperref}  
\hypersetup{colorlinks=true,linkcolor=[rgb]{1.,0.2,0.2},citecolor=[rgb]{0.1,0.4,1.},filecolor=[rgb]{0.7,0.2,0.2},urlcolor=[rgb]{0.7,0.2,0.2}}

\usepackage{color}
\definecolor{blue}{rgb}{0., 0., 1}
\definecolor{lightblue}{rgb}{0.1,0.4,1.}

\newcommand {\CL}{MACS~0416}
\newcommand {\LT}{\texttt{LensTool}}
\newcommand {\T}{Table\,}
\newcommand {\Sec}{Sec.\,}
\newcommand {\Fig}{Fig.\,}
\newcommand {\Eq}{Eq.\,}
\newcommand {\qth}{\texttt{LM-4HALOS}}
\newcommand {\BCGs}{\texttt{LM-BCGs}}
\newcommand {\HBCGs}{\texttt{LM-HLBCGs}}
\newcommand {\shear}{\texttt{LM-SHEAR}}
\newcommand {\IDFB}{$Gal\mbox{-}8971$}

\newcommand {\sn}{$\left<S/N\right>$}

\begin{document} 
\title{A new high-precision strong lensing model of the galaxy cluster MACS~J0416.1$-$2403}
\subtitle{Robust characterization of the cluster mass distribution from VLT/MUSE deep observations}

\author{
P.~Bergamini \inst{\ref{inafbo}} \fnmsep\thanks{E-mail: \href{mailto:pietro.bergamini@inaf.it}{pietro.bergamini@inaf.it}}\fnmsep\thanks{Based on observations collected at the European Southern Observatory for Astronomical research in the Southern Hemisphere under ESO programmes with ID 0100.A-0763(A), 094.A-0115B, 094.A-0525(A).} \and
P.~Rosati \inst{\ref{unife},\ref{inafbo}} \and
E.~Vanzella \inst{\ref{inafbo}} \and
G.~B.~Caminha \inst{\ref{kapteyn},\ref{MPA}} \and
C.~Grillo \inst{\ref{unimi},\ref{darkcc}} \and
A.~Mercurio \inst{\ref{inafna}} \and
M.~Meneghetti \inst{\ref{inafbo}} \and
G.~Angora \inst{\ref{unife},\ref{inafna}} \and
F.~Calura \inst{\ref{inafbo}} \and
M.~Nonino \inst{\ref{inafts}} \and
P.~Tozzi \inst{\ref{arcetri}}
}
\institute{
INAF -- OAS, Osservatorio di Astrofisica e Scienza dello Spazio di Bologna, via Gobetti 93/3, I-40129 Bologna, Italy \label{inafbo} 
\and
Dipartimento di Fisica e Scienze della Terra, Universit\`a degli Studi di Ferrara, via Saragat 1, I-44122 Ferrara, Italy \label{unife}
\and
Kapteyn Astronomical Institute, University of Groningen, Postbus 800, 9700 AV Groningen, The Netherlands \label{kapteyn}
\and
Max-Planck-Institut für Astrophysik, Karl-Schwarzschild-Str. 1, D-85748 Garching, Germany \label{MPA}
\and
Dipartimento di Fisica, Universit\`a  degli Studi di Milano, via Celoria 16, I-20133 Milano, Italy \label{unimi}
\and
Dark Cosmology Centre, Niels Bohr Institute, University of Copenhagen, Jagtvej 128, DK-2200 Copenhagen, Denmark \label{darkcc}
\and
INAF -- Osservatorio Astronomico di Capodimonte, Via Moiariello 16, I-80131 Napoli, Italy \label{inafna}
\and
INAF -- Osservatorio Astronomico di Trieste, via G. B. Tiepolo 11, I-34143, Trieste, Italy \label{inafts}
\and
INAF -- Osservatorio Astrofisico di Arcetri, Largo E. Fermi, I-50125, Firenze, Italy \label{arcetri}
           }

   \date{Received February 14, 2020; accepted February 14, 2020}

% \abstract{}{}{}{}{} 
% 5 {} token are mandatory

  \abstract
  {
    We present a new high-precision parametric strong lensing model of the galaxy cluster MACS~J0416.1$-$2403, at $z=0.396$, which takes advantage of the MUSE Deep Lensed Field (MDLF), with 17.1h integration in the northeast region of the cluster, and Hubble Frontier Fields data. We spectroscopically identify 182 multiple images from 48 background sources at $0.9\!<\!z\!<\!6.2$, and 171 cluster member galaxies. Several multiple images are associated to individual clumps in multiply lensed resolved sources. By defining a new metric, which is sensitive to the gradients of the deflection field, we show that we can accurately reproduce the positions of these star-forming knots despite their vicinity to the model critical lines. The high signal-to-noise ratio of the MDLF spectra enables the measurement of the internal velocity dispersion of 64 cluster galaxies, down to $m_{F160W}=22$. This allowed us to independently estimate the contribution of the subhalo mass component of the lens model from the measured Faber-Jackson scaling relation. Our best reference model, which represents a significant step forward compared to our previous analyses, was selected from a comparative study of different mass parametrizations. The root-mean-square displacement between the observed and model-predicted image positions is only $0.40\arcsec$, which is 33\% smaller than in all previous models. The mass model appears to be particularly well constrained in the MDLF region. We characterize the robustness of the magnification map at varying distances from the model critical lines and the total projected mass profile of the cluster.
  }


   \maketitle
%
%-------------------------------------------------------------------
%@arxiver{Images.pdf,knots.pdf,SI.pdf}

\section{Introduction}

In recent years, strong gravitational lensing has become one of the most effective techniques to characterize the total mass distribution in the inner regions of galaxy clusters, where multiple images of background sources are formed and provide crucial constrains on lens models. The total mass profile of galaxy clusters from kiloparsec up to megaparsec scales can be derived by combining strong lensing with other mass tracers, such as weak lensing \citep[e.g.,][]{Umetsu_2014, Hoekstra_2015,Melchior_2015}, dynamical methods using inner stellar kinematics of the central brightest cluster galaxy \citep[BCG,][]{Sartoris_2020} and the phase space of cluster galaxies \citep{Biviano_2013_macs1206,Stock_2015}, as well as X-ray hydro-static analysis \citep[e.g.,][]{Ettori_2013}. 

By comparing the observed mass density profiles with those predicted by  $N$-body and hydrodynamical simulations, one can test the $\Lambda$ cold dark matter (CDM) paradigm of structure formation \citep[e.g.,][]{Merten_2015}. In particular, a robust characterization of the mass distribution in the cores of galaxy clusters, separating the baryonic and the dark matter (DM) components, can reveal missing physical ingredients in cosmological simulations or possibly constrain physical properties of DM \citep[e.g.,][]{Newman_2013, Grillo_2015,Natarajan_2017,Annunziatella_2017, Bonamigo_2018,Meneghetti_2020}. 

High-precision strong lensing models can be effectively used to study the mass distribution of cluster substructures (or subhalos), particularly when internal kinematics of cluster galaxies is taken into account \citep[see][hereafter B19]{Bergamini_2019}. 
A recent study by \cite{Meneghetti_2020} finds an inconsistency between the number of observed galaxy-galaxy strong lensing (GGSL) systems in massive clusters with those predicted by state-of-the-art cosmological simulations. This study relies on high-precision strong lensing models, such as the one presented here, to dissect the subhalo population  from the total projected mass distribution of clusters. 

In addition to total mass distributions, magnification maps obtained from strong lensing models of galaxy clusters are fundamental tools to study the astrophysical properties of lensed and highly magnified background sources. Specifically, the lensing magnification allows one to resolve details on scales of a few tens of parsecs at redshift 2-6 \citep[see][]{vanz_paving,Vanzella_ID14,johnson17,rigby17,cava18,vanz19,Vanzella_2020}. Thus, until the advent of new observational facilities, such as the James Webb Space Telescope (JWST) and the Extremely Large Telescope (ELT), strong lensing represents the only available technique to characterize the high redshift population of faint sources, that is the progenitors of present day galaxies and globular clusters, which are expected to play an important role in the re-ionization process of the Universe \citep[e.g.,][]{Boylan_2018,ma20_gc_eor,he20_ricotti,vanz_sunburst}.

Over the last decade, we have witnessed important progress in cluster lens models thanks to several observational campaigns that have provided high-quality photometric and spectroscopic data on a sizable sample of massive galaxy clusters, which are generally powerful gravitational lenses. The first of these campaigns, the Cluster Lensing And Supernova survey with Hubble \citep[CLASH,][]{Postman_2012_clash} provided panchromatic data on 25 massive clusters with the WFC3 and ACS cameras on Hubble Space Telescope (HST). The Re-ionization Lensing Cluster Survey \citep[RELICS,][]{Coe_2019} extended this HST study to 41 clusters. Significantly deeper HST observations, in seven ACS/WFC3 bands, were carried as part of the Hubble Frontier Field program \citep[HFF,][]{Lotz_2014HFF, Lotz_2017HFF} on six clusters selected to be powerful lenses.
One of the HFF targets, MACS~J0416.1$-$2403 (hereafter \CL), is the focus of this work. This is a massive X-ray luminous galaxy cluster at $z=0.396$ \citep[][]{Ebeling_MACS}, with a weak lensing mass of $M_{200\mathrm{c}} = (1.04\pm0.22) \times 10^{15} \mathrm{M_{\odot}}$ \citep{Umetsu_2014}.

In addition to the HST imaging campaigns,  spectroscopic follow-up programs were carried out on a subsample of clusters. The CLASH-VLT Large Programme (Rosati et al. in prep.), for example, collected $\sim\!30000$ redshifts in the fields of 13 CLASH clusters with the VIsible Multi-Object Spectrograph (VIMOS) on the ESO's VLT. Approximately two hundreds lensed background sources and multiple images were also identified. The CLASH-VLT spectroscopic campaign on \CL\ was presented in \cite{Balestra_2016}. The ability to identify large number of multiple images in the core of galaxy clusters improved dramatically with the advent of the Multi-Unit Spectroscopic Explorer (MUSE) on the VLT (\citealt{Bacon_MUSE}, see e.g., \citealt{Caminha_macs0416,Lagattuta_2017}).

The accuracy of strong lensing modeling relies critically on the number of spectroscopically confirmed multiple images \citep{Grillo_2015,Johnson_2014,Caminha_2019}.  In fact, spectroscopic information reduces the risk of multiple image misidentifications and breaks the degeneracy between source distances and the total mass of the lens. 
In addition, integral field spectroscopy allows the identification of a complete and pure sample of cluster galaxies which are another crucial ingredient for lens models. 
As demonstrated in B19, MUSE data can also be used to measure the inner stellar kinematics of numerous cluster member galaxies, which can be incorporated into lens models, thus adding an important constraint on the mass of the subhalo population \citep[see also][]{Monna_2017}. 

The dynamical analysis of \CL\ by \cite{Balestra_2016}, using $\sim\! 900$ spectroscopic members, combined with the Chandra X-ray and VLA radio observations, revealed a complex, mostly bi-modal mass distribution, which is likely the result of a pre-merging phase. 
The 30 spectroscopically confirmed multiple images identified in the CLASH-VLT campaign were used in the lens model of \cite{Grillo_2015}, thus improving on previous photometric studies \citep{Zitrin_2013}. 
A free-form lens model was presented by \cite{Hoag_2016}, using an extended spectroscopic coverage of \CL\ based on the Grism Lens-Amplified Survey from Space \citep[GLASS,][]{Treu_2015,Schmidt_2014}. A significant step forward was then made by \cite{Caminha_macs0416} (hereafter C17) by exploiting the first MUSE observations of \CL. This study presented a high-precision lens model based on a new large sample of 102 spectroscopic multiple images (from 37 background sources) and an extended catalog of cluster galaxies (including 144 spectroscopic members). 
In \cite{Bonamigo_2018}, we refined the C17 lens model by including the mass component associated to the hot-gas, which is traced by deep Chandra X-ray observations \citep{Ogrean_2015} and dominates the cluster baryonic content. 

In this paper, we further improve the lens model of \CL\ by exploiting additional MUSE observations in the northeast part of the cluster, the MUSE Deep Lens Field (MDLF), which is presented in a accompanying paper by \cite{Vanzella_2020}. 
When compared to our previous models (C17 and B19), the combination of the MDLF and the HFF data lead us to identify $\sim\! 80\%$ more multiple images, $\sim\! 20\%$ more spectroscopic galaxies and to measure the inner velocity dispersion of cluster members one magnitude fainter.\footnote{When this paper was close to be submitted, \cite{Richard_2020} presented the results from a lens model of \CL\ based on the new MUSE deep observations. A comparison between the two models will be possible when their model details will be made public.}

The paper is organized as follows. In \Sec\ref{sec:Data}, we present the \CL\ dataset used to develop our new lens model. \Sec\ref{sec:lens_model} describes in detail the mass parametrization of a selected number of lens models. 
In \Sec\ref{sec:res}, we discuss the results of the optimization of four lens models and the criteria leading to the selection of the best reference model. The latter is used to study the robustness of the magnification values of multiple images and to characterize the projected total mass profile of \CL. The main conclusions of our work are drawn in \Sec\ref{sec:conclusions}.

Throughout this work, we adopt a flat $\Lambda$CDM cosmology
with $\Omega_m = 0.3$ and $H_0= 70\,\mathrm{km\,s^{-1}\,Mpc^{-1}}$. Using this cosmology, a projected distance of $1\arcsec$ corresponds to a physical scale of 5.34 kpc at the \CL\ redshift of $z=0.396$. All magnitudes are given in the AB system.

%--------------------------------------------------------------------

\begin{figure}[ht!]
	\centering
	\includegraphics[width=\linewidth]{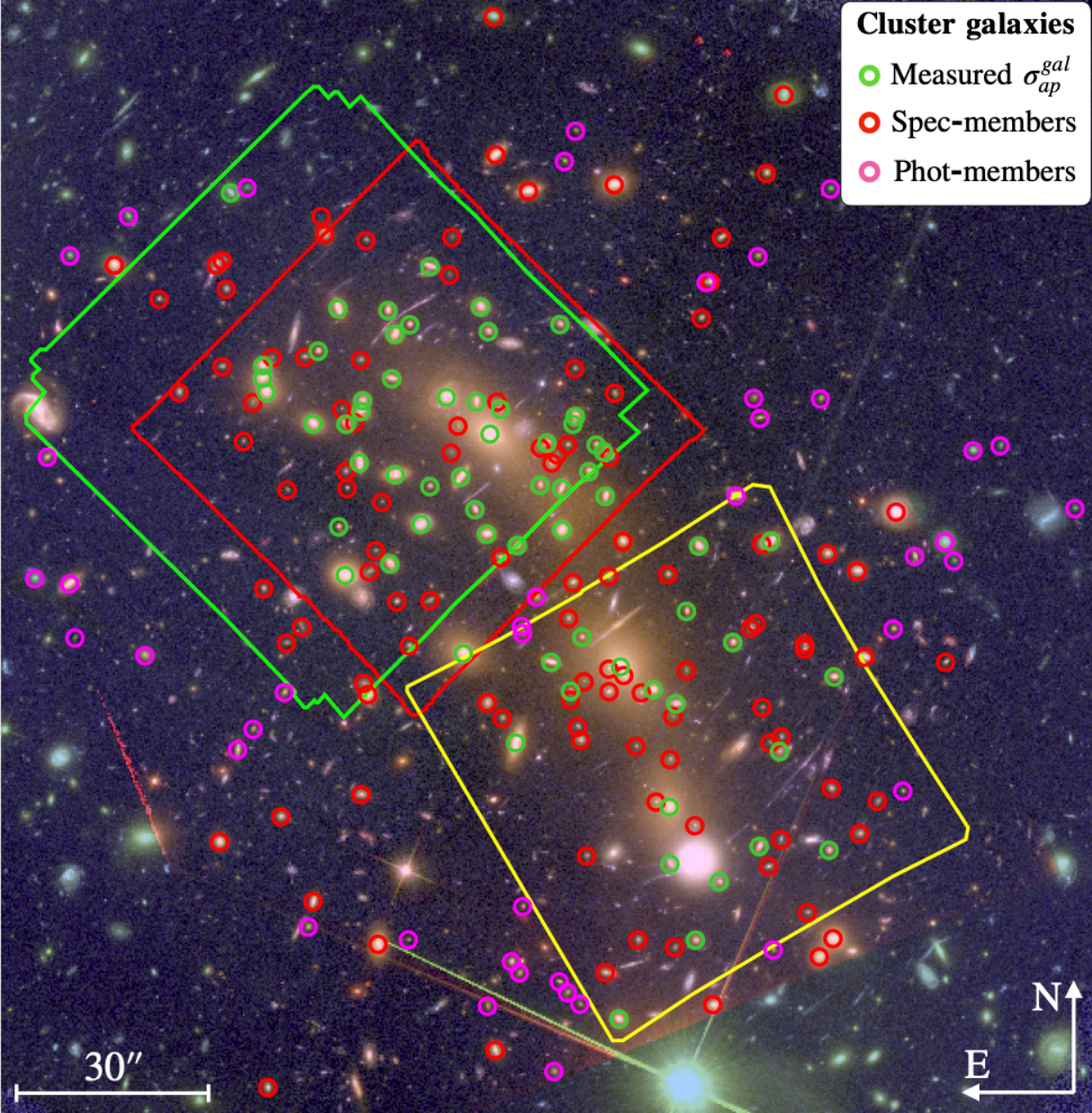}
	\caption{RGB image (F105W+F125W+F140W+F160W,  F606W+F814W, F435W) of \CL\ showing the footprints of the northeast and southwest MUSE pointings in red and yellow, respectively. The new deep MUSE observation in NE region, totaling to 17.1h, is shown in green. Circles mark the 213 cluster galaxies included in our lens models. Red and green circles refer to the 171 spectroscopic members. Green circles correspond to the 64 cluster galaxies for which we measure a reliable internal velocity dispersion from MUSE spectra. The remaining 42 photometric members are encircled in magenta.}
	\label{fig:CM}
\end{figure}

\section{Data}
\label{sec:Data}
In this section, we briefly describe the \CL\ data-set presented in \cite{Vanzella_2020}, specifically the new catalogs of multiple images and cluster member galaxies used in our lens models.

\subsection{HST photometric and MUSE spectroscopic observations of \CL}

This work is based on the HST multi-band imaging data from the CLASH survey
and the HFF program. 
\CL\ has been the target of several spectroscopic follow-up campaigns. The redshift measurements over a wide ($\sim\! 20$ arcmin across) field were obtained as part of the CLASH-VLT program with VLT/VIMOS and presented in \citet{Balestra_2016}. \citet{Grillo_2015} constructed a lens model using 30 spectroscopically confirmed multiple images from this initial spectroscopic dataset. The number of multiple images with a spectroscopic confirmation increased by more than a factor of three with the advent of MUSE observations, as described by C17. MUSE data-cubes have a field-of-view of $1\,\mathrm{arcmin}^2$ that is spatially sampled with $0.2\arcsec\times0.2\arcsec$ pixels. The wavelength range extends from $4700\,\AA$ to $9350\,\AA$ with a dispersion of $1.25\,\AA/$pix, and a spectral resolution of $\sim2.6\,\AA$ fairly constant across the entire spectral range. 

The study of C17 included two MUSE pointings, one to the northeast (NE) of 2h exposure ($0.6\arcsec$ seeing) and one to the southwest (SW) of 11h integration ($1.0\arcsec$ seeing), the latter however provided spectra with a signal-to-noise ratio below expectations (see comment in that paper). 
In this work, we take advantage of the MDLF dataset, which extended the integration time in most of the NE field to a total of 17.1h, with a final seeing of $0.6\arcsec$ \citep{Vanzella_2020}.\footnote{programs IDs: NE (2h): GTO 094.A-0115B (P.I. J. Richard), SW (11h): 094.A0525(A) (P.I. F.E. Bauer), NE-deep (15.1h): 0100.A-0763(A) (P.I. E. Vanzella)} 
In \Fig\ref{fig:CM}, we show the footprints of the three MUSE pointings overlaid onto the HST/RGB image of the cluster.

\begin{figure*}[h!]
	\centering
	\includegraphics[width=0.832\linewidth]{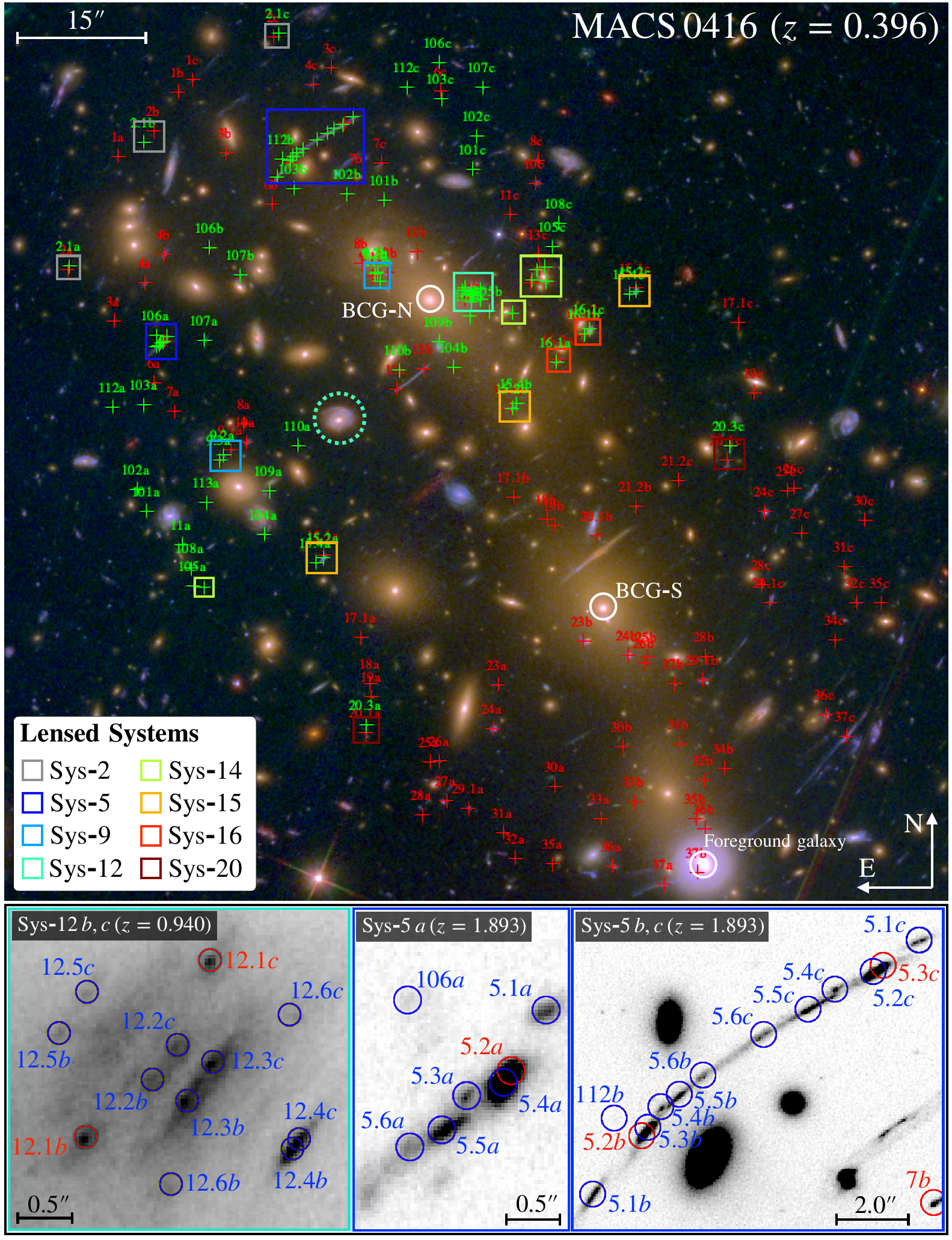}
	\caption{HST RGB image (F814W, F606W, F435W) of the central region of the galaxy cluster \CL\ at z=0.396. The colored crosses mark the position of the 182 spectroscopic multiple images used to constrain the lens models: red crosses mark the 100 images in common with the catalog by C17, while the newly identified images are in green. The two BCGs of the cluster and the foreground galaxy at $z=0.112$ are encircled in white. Colored rectangles highlight the systems of multiple images produced by eight background sources resolved into multiple clumps (see \Sec\ref{sec:image_cat}). The dotted cyan circle marks the position of the third predicted image of Sys-12. The bottom panels show three zoom-in images of systems 12 and 5, obtained from a median stack of the F814W, F606W, and F435W HST filters. The red and blue circles correspond to the red and green crosses in the main panel.}
	\label{fig:cluster}
\end{figure*}

\subsection{Catalog of multiple images}
\label{sec:image_cat}
 The latest public catalog of spectroscopically confirmed multiple images in MACS0416, before this work, was released by C17. It included a total of 102 images from 37 background sources. The same catalog was used in the lens models presented by \cite{Bonamigo_2018} and B19. As described in \cite{Vanzella_2020}, the combined analysis of the MDLF observations and the CLASH/HFF multi-band images has led to the identification of 182 secure multiple images, at $0.9<z<6.2$, which are used in our new lens model.

We assign to each multiple image an ID containing a number and a letter so that images with the same number but different letters belong to the same family of multiple images. In \CL, we find eight resolved lensed sources, each containing two or more point-like knots (or clumps) at the same redshift, which are often embedded into a diffuse light emission. Multiple images of the same clump form a family, while we call the set of images coming from the same resolved source, a ``system'' (abbreviated Sys) of multiple images. Thus, the 182 multiple images are found associated to 66 independent families, drawn from 48 different background sources. Images belonging to a system are characterized by an identical integer part of their ID number, but by a different fractional part. Our analysis demonstrates that these multiply imaged clumps are very efficient in constraining the position of the critical lines of the lens models (e.g., see Sys-12 discussed below). 

Two examples of systems of multiple images are given in the cut-outs of \Fig\ref{fig:cluster}. Sys-12 is made by six image families, with IDs from 12.1 to 12.6, containing two multiple images ($b,\,c$) of the same clumps at $z=0.940$. Instead, Sys-5 is formed by six families (from 5.1 to 5.6) of three multiple images each ($a,\,b,\,c$). These resolved sources are indicated with rectangles in \Fig\ref{fig:cluster}, with a different color highlighting each of the eight systems. Red crosses correspond to multiple images belonging the previous catalog by C17, while green crosses are the newly discovered images. We note that two images of the same background source at $z=3.923$ (identified as 22b and 22c in the C17) have been excluded from our final catalog since the elongation of their Ly$\alpha$ emission does not allow a precise determination of their positions. 
The northern region of the cluster includes 80 out of 82 of the new multiple images, where the deepest MUSE pointing was carried out. The remaining two images, identified as 20.3a and 20.3c, are new clumps identified in the source at $z=3.222$ in the SW field. 

\begin{figure}[ht!]
	\centering
	\includegraphics[width=\linewidth]{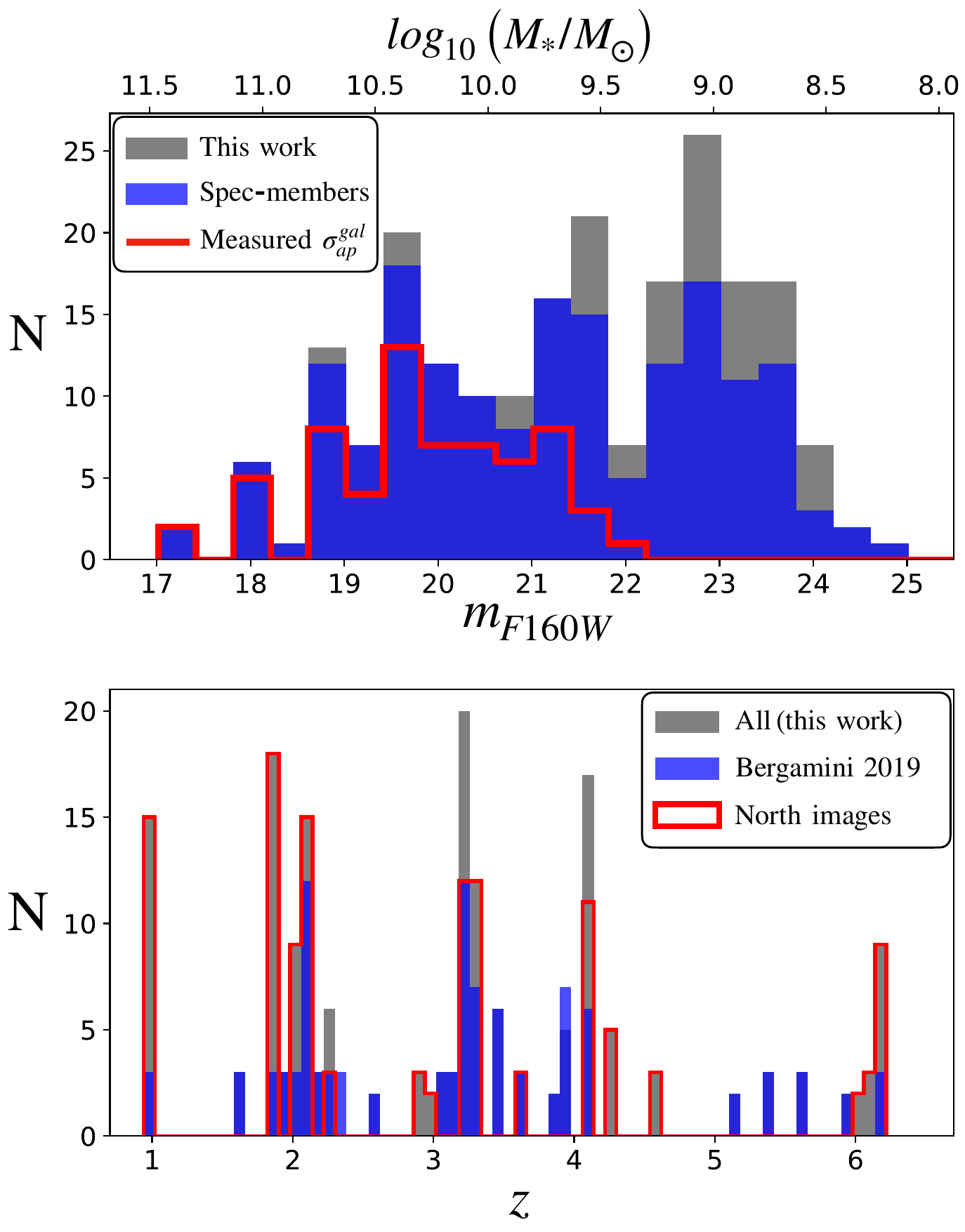}\caption{
	{\it Top:} Distribution of cluster member galaxies as a function of their magnitudes in the HST/F160W filter. The new sample of cluster members used in our lens models is plotted in gray, with the spectroscopic members indicated in blue. Cluster members with a reliable measurement of their internal velocity dispersion are in red. The stellar mass on the top axis is based on  the empirical relation in \citealt{Grillo_2015}.
	{\it Bottom:} Redshift distribution of the 182  multiple images used to constrain the lens models (gray). The previous sample of multiple images from C17 is shown in blue. The red histogram refers to only the multiple images from the deep NE field (above the white dashed line in \Fig\ref{fig:DM_images}).
	 }
	\label{fig:histos}
\end{figure}

In \Fig\ref{fig:histos} (lower panel), we show in gray the redshift distribution of the new set of 182 multiple images, ranging from $z=0.940$ to $z=6.145$, while the blue histogram refers to the previous sample published in C17. The 125 images in the deep NE region are drawn in red. 

\subsection{Cluster members selection and measured stellar velocity dispersions}
\label{sec: data_CM}
Exploiting the high signal-to-noise ratio of the cluster member spectra extracted from the MDLF pointing in the NE region of \CL, we extend the publicly released cluster member catalog used in the lens model of C17. 
The latter contained 144 spectroscopically confirmed cluster members, which are defined as those galaxies in the HST/WFC3 field-of-view, brighter than $m_{F160W}=24$, and with a velocity within $\pm3000\,\mathrm{km\,s^{-1}}$ in cluster rest frame centered at $z=0.396$ (this corresponds to the redshift range $[0.382 \mbox{-} 0.410]$). To maximize sample completeness, this sample was complemented with 49 photometric members selected from their color space distribution based on 12 CLASH photometric bands,  with $m_{F160W}<24$, as described in \citealt{Grillo_2015}. 

The revised catalog of cluster members, to be used for the new lens models, includes 19 additional spectroscopic members based on MDLF observations. Four of these galaxies are fainter than $m_{F160W}=24$, they are however included in the sample due to their secure membership. We also add to this catalog seven extra spectroscopic members based on new HFF/F160W photometry, which is needed for the lens model. 
Moreover, we add to the final sample a bright galaxy ($m_{F160W}$=18.11) at $z=0.4111$ whose peculiar velocity is only $245\,\mathrm{km\,s^{-1}}$ above the upper bound of the velocity range chosen for the cluster member selection. Due to its high F160W luminosity, we expect this galaxy to have a non-negligible impact on the cluster mass model. We remove a faint cluster member (ID $Gal\mbox{-}739$ in C17) since it is below the magnitude limit based on revised photometry. 

Summarizing, our final cluster member catalog counts a total of 213 galaxies, 171 have a secure spectroscopic redshift, while 42 members are still photometric members. We note that a recent method based on a convolution neural network technique, which was developed to identify cluster members using multi-band HST image cutouts \citep{Angora_2020}, confirms 40 out of the 42 photometric members included in this work.

Following the study by B19, we measure the line-of-sight stellar velocity dispersion, $\sigma_{ap}^{gal}$, of a large number of cluster galaxies by taking advantage of the increased mean signal-to-noise ratio ($\left<S/N\right>$) of galaxy spectra in the MDLF (a factor $\sim\! 2$). We use spectral extraction apertures of  $0.8\arcsec$ radius, which yields the best compromise between high \sn\ and low contamination from nearby bright sources. 
Velocity dispersions are measured by cross-correlating the observed galaxy spectra with a set of stellar templates, using the  02/2018 version of the public software pPXF \citep{Cappellari_2004,Cappellari_2017}, as discussed in B19.
pPXF fits are performed over the wavelength range $5000\mbox{-}7260\,\AA$ to avoid the red part of the MUSE spectra, which is contaminated by skyline residuals, and at the same time includes all relevant galaxy absorption lines. In addition to regions containing sky emission lines, masking is applied to the wavelength range $5800\mbox{-}6000\,\AA$, which is highly contaminated by the sodium laser emission of the adaptive optics system used in most of the MDLF observations.  

\begin{figure}[t!]
	\centering
	\includegraphics[width=\linewidth]{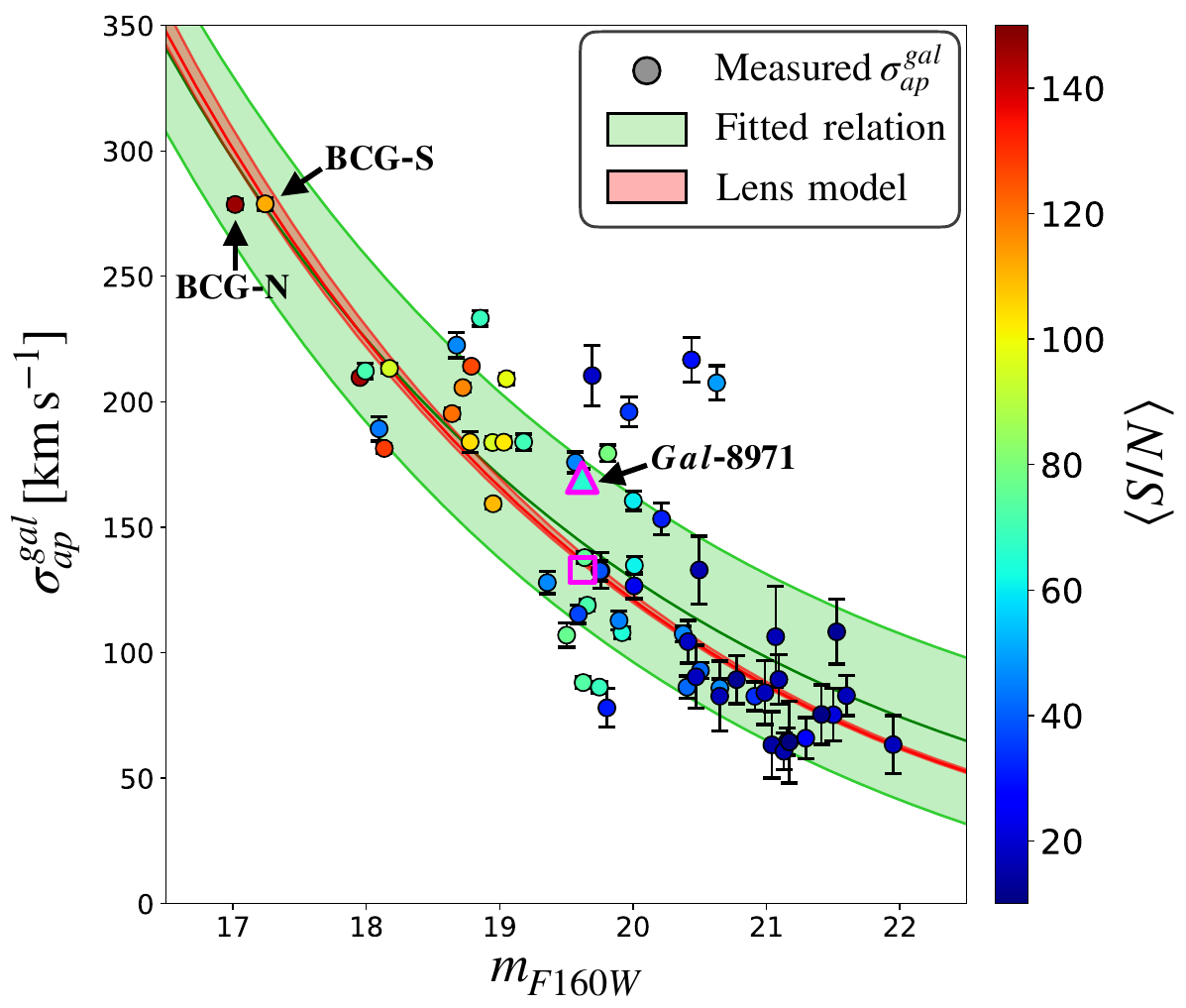}
	\caption{Measured internal stellar velocity dispersions of 64 cluster member galaxies as a function of their magnitudes in the HST/F160W filter (filled circles). Their colors encode the mean signal-to-noise ratio of galaxy spectra (\sn).
	The magenta triangle refers to the bright galaxy member in Sys-14 ($Gal\mbox{-}8971$).
	The green solid line is the best-fit $\sigma_{ap}^{gal}\,\mbox{-}\,m_{F160W}$ relation obtained as described in \Sec\ref{sec: galaxy_mass_distribution}, while the light green area corresponds to measured mean scatter around the best-fit ($\Delta\sigma_{ap}$). The red band corresponds to the 68\% confidence level of the $\sigma\,\mbox{-}\,m_{F160W}$ relation obtained from the optimization of our reference lens model (\qth). The magenta square indicates the velocity dispersion of $Gal\mbox{-}8971$ and its $1\mbox{-}\sigma$ error, as predicted by the lens model (see \Sec\ref{sec:lens_model}).
	 }
	\label{fig:green_plots}
\end{figure}

As explained in B19, to ensure robust velocity dispersion measurements we limit our sample of galaxies with measured stellar kinematics to $\left<S/N\right>>10$ and $\sigma_{ap}^{gal}>60\,\mathrm{km\,s^{-1}}$. We also apply small corrections to  $\sigma_{ap}^{gal}$ and its uncertainties, $d\sigma_{ap}^{gal}$, based on spectral simulations (see appendix A in B19).  The final sample of  cluster members with internal kinematics includes 64 galaxies. We note that the MDLF pointing allows us to measure reliable velocity dispersions for 15 additional galaxies down to $m_{F160W}\sim22$, that is approximately a magnitude fainter than the limit adopted in B19. The sample of cluster members with measured internal kinematics is indicated in \Fig\ref{fig:CM} (green circles). In \Fig\ref{fig:green_plots}, we show the measured $\sigma_{ap}^{gal}$ values as a function of magnitude in the F160W filter.

%--------------------------------------------------------------------
\section{Strong lensing models}
\label{sec:lens_model}

To model the mass distribution of \CL\ we use the public software \LT\ \citep{Kneib_lenstool,Jullo_lenstool,Jullo_Kneib_lenstool} for strong lensing modeling. Following the method adopted in our previous studies, we decompose the cluster total mass into several mass components, based on simple parametric models. A set of free-parameters ($\pmb{\xi}$) is determined by constraining the mass models with the positions of the 182 multiple images (\Sec\ref{sec:image_cat}), $\mathbf{x}^{obs}$, thereby maximizing the following posterior probability distribution function \citep[see][]{Caminha_rxc2248}:

\begin{equation}
    \label{eq.: posterior_lt}
    P\left(\pmb{\xi}|\mathbf{x}^{obs}\right) \propto P\left(\mathbf{x}^{obs}|\pmb{\xi}\right) \times P\left(\pmb{\xi}\right)\pmb{,}
\end{equation}

\noindent where $P\left(\pmb{\xi}\right)$ correspond to prior probability distributions for the model free-parameters, while the likelihood $P\left(\pmb{\xi}|\mathbf{x}^{obs}\right)$ is given by:

\begin{equation}
    \label{eq.: likelihood_lt}
   {\cal L} \equiv P\left(\mathbf{x}^{obs}|\pmb{\xi}\right) \propto \exp \left(-\frac{1}{2}\chi^2\left(\pmb{\xi}\right)\right).
\end{equation}

\noindent The lens model $\chi^2$ quantifies the displacement  on the lens plane between the observed and model-predicted positions of the multiple images ($\mathbf{x}^{pred}$), given the set of model parameters $\pmb{\xi}$. To compute the $\chi^2$ value an isotropic uncertainty, $\Delta{x}_{i,j}$, on the observed positions of the images is assumed \citep{Jullo_lenstool}:

\begin{equation}
    \label{eq.: chi_lt}
    \chi^2(\pmb{\xi}) := \sum_{j=1}^{N_{fam}} \sum_{i=1}^{N_{im}^j} \left(\frac{\left\| \mathbf{x}_{i,j}^{obs} - \mathbf{x}_{i,j}^{pred} (\pmb{\xi}) \right\|}{\Delta x_{i,j}}\right)^2.
\end{equation}

\noindent The index $i$ runs over the multiple images belonging to the same $j$-th family (e.g., 1a, 1b, 1c for system 1), while the index $j$ runs over all families (in our case $N_{fam}=66$). So $N_{im}^j$ is the number of observed multiple images coming from the same $j$-th family (e.g., $N_{im}^{j=1}=3$).

Since the lens model is constrained by the position of $N_{im}^{tot}=\sum_{j=1}^{N_{fam}} N_{im}^j$ observed multiple images, by defining $N_{par}$ as the total number of model free-parameters, we can write the number of degree-of-freedom (DoF) of the lens model as:

\begin{equation}
    \label{eq.: DoF}
    \mathrm{DoF} = 2 \times N_{im}^{tot} - 2 \times N_{fam} - N_{par} = N_{con}-N_{par}.
\end{equation}

The term $2 \times N_{fam}$ stems from the fact that the unknown positions of the $N_{fam}$ background sources are additional free parameters of the model. Thus, $N_{con}$ is the effective number of available constraints.

To sample the lens model posterior distribution in Eq.\ref{eq.: likelihood_lt}, \LT\ exploits a Bayesian Markov chain Monte Carlo (MCMC) technique. After the removal of a large burn-in phase ($\sim7\times10^3$ samples, obtained using ten walkers), the final MCMC chains of our lens models contain at least $10^5$ samples of free parameters' values. The uncertainties on the model free-parameters are determined by re-sampling the posterior distribution once the error on the positions of the observed multiple images ($\Delta x_{i,j}$) is re-scaled so that the reduced $\chi^2$ is close to one (the initial value $\Delta x_{i,j}$ is set to 0.5\arcsec (1.0\arcsec) for HST (MUSE) detected images). 

To quantify the goodness of our lens models, we use three main indicators. The first, as customary, is the root-mean-square separation between the observed and model-predicted positions of multiple images, $\Delta_{rms}$ \citep[see e.g.,][]{Caminha_rxc2248}:  

\begin{equation}
    \label{eq.: rms_lt}
    \Delta_{rms}=\sqrt{\frac{1}{N_{im}^{tot}}\sum_{i=1}^{N_{im}^{tot}}\left\|\boldsymbol{\Delta_i} \right\|^2} ,\quad \mathrm{with}\quad \boldsymbol{\Delta_i} = \mathbf{x}_{i}^{obs} - \mathbf{x}_{i}^{pred},
\end{equation}

\noindent where $\boldsymbol{\Delta_i}$ is the displacement between the $i$-th observed and predicted image. The second and the third indicators are the Bayesian information criterion (BIC, \citealt{Schwarz_1978}), and the Akaike information criterion (AIC, \citealt{Akaike_1974}) defined as:
\begin{flalign}
\begin{aligned}
\label{eq: BICAIK}
        &\mathrm{BIC}\equiv -2 \ln({\cal L}_{max}) + N_{par} \ln(N_{con}), \\ 
        &\mathrm{AIC}\equiv -2 \ln({\cal L}_{max}) + 2\,N_{par}.
\end{aligned}
\end{flalign}
\noindent ${\cal L}_{max}$ is the maximum value of the likelihood in \Eq\ref{eq.: likelihood_lt}.

These information criteria ensure that the introduction of extra free parameters in a lens model is justified by a corresponding increase of the model likelihood, thus avoiding over-fitting. As a general rule of thumb, the best cluster lens model has the lowest $\Delta_{rms}$, BIC and AIC values.

\begin{table*}[h!]
	\tiny
	\def\arraystretch{1.6}
	\centering    
	\begin{tabular}{|c|c|c|c|c|c|}
	   \hline
	   \multicolumn{6}{|c|}{\textbf{Measured parameters of the scaling relations}}\\
	   \hline
	   \boldmath{$N(\sigma_{ap}^{gal})$} & \boldmath{$m_{F160W}^{ref}$} & \boldmath{$\sigma_{ap}^{ref}\ \mathrm{[km\ s^{-1}]}$} & \boldmath{$\alpha$} & \boldmath{$\Delta\sigma_{ap}\ \mathrm{[km\ s^{-1}]}$} & \boldmath{$\beta_{cut}(\gamma=0.2)$} \cr
	   \hline
	   64 & 17.02 & $295.2_{-16.6}^{+17.3}$ & $0.30_{-0.03}^{+0.03}$  & $33.1_{-2.9}^{+3.4}$ & $0.60_{-0.06}^{+0.06}$ \cr
	   \hline
	\end{tabular}
	\smallskip
    \caption{Main parameters of the $\sigma_{ap}^{gal}\mbox{-}m_{F160W}$ and $r_{cut}^{gal}\mbox{-}m_{F160W}$ scaling relations obtained from the measured stellar velocity dispersions of $N(\sigma_{ap}^{gal})=64$ cluster member galaxies. For each parameter, we quote the median value and the 16-th, 84-th percentiles of its marginalized posterior distribution (see \Fig\ref{fig:deg_scaling}). The normalization $\sigma_{ap}^{ref}$ is computed at the reference magnitude $m_{F160W}^{ref}=17.02$ of the northern BCG. The slope $\beta_{cut}$ of the $r_{cut}^{gal}\mbox{-}m_{F160W}$ relation is inferred from \Eq\ref{eq.: slopes} (see text).  
    }
	\label{table:kinemacs_scaling_relations} 

\end{table*}

\subsection{Mass components}
\label{sec:mass_comp}
The overall total mass distribution of \CL\ (or equally its total gravitational potential, $\phi_{tot}$) is divided into the following sum of parametric mass profiles:

\begin{equation}
    \label{eq.: pot_dec}
    \phi_{tot}= \sum_{i=1}^{N_h}\phi_i^{halo}+\sum_{j=1}^{N_g}\phi_j^{gal}+\phi_{foreg}+\phi_{\kappa,\gamma}.
\end{equation}

\noindent The first term takes into account the $N_h$ cluster-scale smooth halos of the cluster potential ($\phi_{i}^{halo}$), while the second one is associated to the $N_g$ cluster member galaxies (or subhalos), each with gravitational potentials $\phi_j^{gal}$. The third term, $\phi_{foreg}$, is the contribution from a prominent foreground galaxy residing in the SW region of the cluster. The last term, $\phi_{\kappa,\gamma}$, refers to a possible constant convergence, or shear, associated to extra mass unaccounted for in the cluster field.

In our study, we have explored a large number of lens models with different mass parametrizations and number of subcomponents. In the following two subsections, we describe the best four lens models selected on the basis of the aforementioned criteria, which are referred to as \qth, \BCGs, \HBCGs, and \shear. In \Sec\ref{sec:res}, we explain why the \qth\ model is finally adopted as our reference model.

\subsection{Cluster-scale mass distribution}
\label{sec: clusterscale_mass_distribution}
Most of the total mass of galaxy clusters is in the form of smooth halos that extend over a scale of hundreds to thousands of kiloparsec. These cluster-scale halos are dominated by the DM component, with a non-negligible fraction of hot-gas and stars responsible for the intra-cluster light (ICL) emission.

The cluster-scale component of our \LT\ models 
is parametrized as a sum of elliptical dual pseudo-isothermal mass distributions \citep[dPIE,][]{Limousin_lenstool, Eliasdottir_lenstool, Bergamini_2019}. The set of free-parameters includes: two parameters for their sky position, two for the ellipticity $e=\frac{a^2-b^2}{a^2+b^2}$ (where $a$ and $b$ are the semi-major and semi-minor axis of the dPIE profile) and the position angle ($\theta$) measured counterclockwise from the west direction, additional three parameters are the central velocity dispersion $\sigma_0$, the core radius $r_{core}$, and the truncation radius $r_{cut}$. 

All the lens models we have developed contain four cluster-scale dPIEs that are used to parameterize the hot-gas mass content, obtained by fitting the Chandra deep X-ray surface brightness distribution \citep[see][]{Bonamigo_2018}. These are fixed profiles and therefore do not contribute to the number of free-parameters. In the \BCGs, \HBCGs, and \shear\ lens models, the DM and ICL content of the cluster halo is parametrized using three dPIEs of infinite truncation radius. Two elliptical dPIEs have center positions close to the cluster BCGs, while a third circular dPIE is left free to move in the NE region around a minor galaxy over-density (as in B19 and C17). 
In the \qth\ model, we add a fourth cluster-scale elliptical dPIE whose position is left free to vary in the southern region of the cluster. This extra halo significantly reduces the model $\Delta_{rms}$, particularly around the BCG-S. The \shear\ model was also studied to test the impact of possible undetected massive structures in the cluster outskirts (i.e.,  $\phi_{\kappa,\gamma}\ne 0$ in \Eq\ref{eq.: pot_dec}). In this model, the $x$ and $y$ shear components ($\gamma_x$ and $\gamma_y$), and the value of the convergence ($\kappa$) are additional free-parameters.

\begin{figure}[t!]
	\centering
	\includegraphics[width=\linewidth]{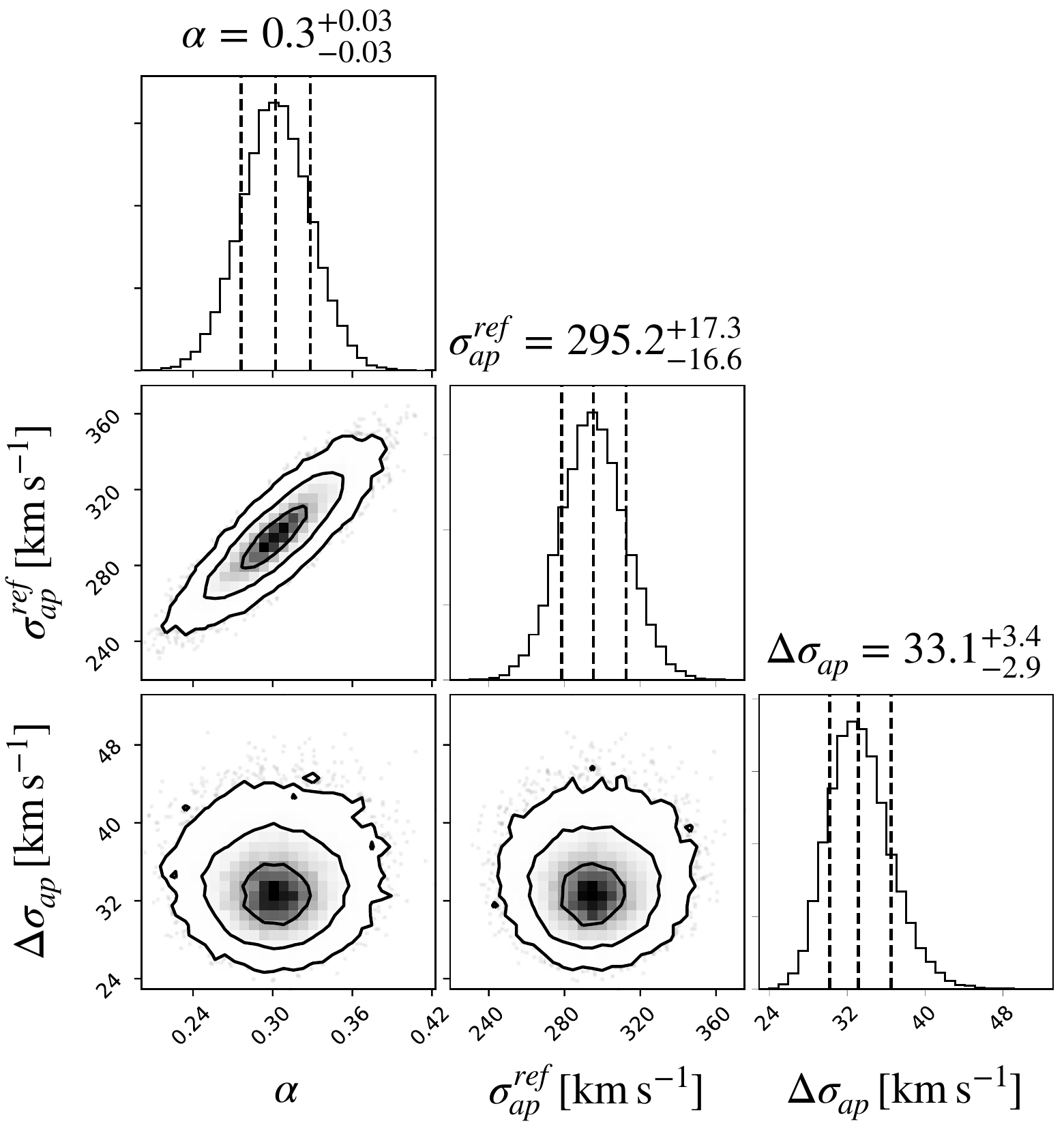}
	\caption{
	Marginalized posterior distributions for the fitting parameters of the $\sigma_{ap}^{gal}\mbox{-}m_{F160W}$ scaling relation (see \Fig\ref{fig:green_plots}). The normalization $\sigma_{ap}^{ref}$ is computed at the magnitude of the BCG-N, $\alpha$ is the value of slope of the scaling relation, $\Delta\sigma_{ap}$ quantifies the mean scatter of the $\sigma_{ap}^{gal}$ values around the best-fit scaling relation. The median values and [16-th, 84-th] percentiles of the marginalized posterior distributions are quoted in the titles.
	}
	\label{fig:deg_scaling}
\end{figure}

\subsection{Galaxy-scale mass distribution}
\label{sec: galaxy_mass_distribution}
Here we describe how we model the total mass (DM plus baryons) content of cluster member galaxies.
Each subhalo is parametrized as a circular dPIE profile, with negligible core radius, whose position is centered on the peak of the stellar light emission. As customary, to reduce the number of free-parameters, we adopt the following scaling relations for the central velocity dispersions ($\sigma_0^{gal}$) and truncation radii ($r_{cut}^{gal}$), as a function of galaxy luminosity \citep{Jorgensen_96,Natarajan_1997}:

\begin{equation}
    \sigma^{gal}_{LT,i}= \sigma^{ref}_{LT} \left(  \frac{L_i}{L_{ref}} \right)^{\alpha},
    \label{eq.: Scaling_relation_sigma}
    \end{equation}
    \begin{equation}
    r^{gal}_{cut,i}= r^{ref}_{cut} \left(  \frac{L_i}{L_{ref}} \right)^{\beta_{cut}}.
    \label{eq.: Scaling_relation_rcut}
\end{equation}

\noindent In \Eq\ref{eq.: Scaling_relation_sigma}, we introduce the \LT\ fiducial velocity dispersion $\sigma_{LT}$ that is related to the central velocity dispersion of the dPIE by $\sigma_0=\sqrt{\frac{3}{2}}\sigma_{LT}$. The two normalizations $\sigma_{LT}^{ref}$ and $r^{ref}_{cut}$ are computed at the reference luminosity $L_{ref}$. For the $L_{i}$ luminosities, we use F160W Kron magnitudes of cluster galaxies, as a good proxy of their total mass \citep{Grillo_2015}. The BCG-N magnitude, $mag_{F160W}^{ref}=17.02$, is used for the values of $L_{ref}$ in Eqs.\,\ref{eq.: Scaling_relation_sigma} and \ref{eq.: Scaling_relation_rcut}.
A third scaling relation with slope $\beta_{core}=0.5$ is implemented in \LT\ to scale the dPIE core radius. However, since we assume a negligible value for the reference core radius ($r_{core}^{ref}=1\arcsec\times10^{-4}=0.5\,\mathrm{pc}$), the very inner region of the galactic dPIEs is well approximated by a singular isothermal behavior.

Based on the Bayesian technique described in B19 \footnote{We exploit the python implementation of the Affine-Invariant MCMC Ensemble sampler \citep[][https://emcee.readthedocs.io/en/latest/]{gw2010,emcee_2013}.}, we fit the slope $\alpha$ and the normalization $\sigma_{ap}^{ref}$ of the $\sigma_{ap}^{gal}\,\mbox{-}\,m_{F160W}$ relation, that is the Faber-Jackson relation in \Eq\ref{eq.: Scaling_relation_sigma} \citep{Faber-Jackson_1976}, using the 64 stellar velocity dispersions $\sigma_{ap}^{gal}$ (see \Sec\ref{sec: data_CM}). As in B19 (appendix B), we consider 100 walkers performing 5000 steps each, with the following priors $\left[\sigma_{min}^{ref},\,\sigma_{max}^{ref}\right]=\left[100,\,600\right]$, $\left[\alpha_{min},\,\alpha_{max}\right]=\left[0.0,\,0.5\right]$, and $\left[(\Delta\sigma_{ap})_{min},\,(\Delta\sigma_{ap})_{min}\right]=\left[0,\,100\right]$. A burn-in phase equal to two times the auto-correlation time of each parameter (about 77 steps) is removed from the final chain. The solid green line in \Fig\ref{fig:green_plots} corresponds to the best-fit scaling relation, while the light-green band represents the mean scatter ($\Delta\sigma_{ap}$) of the relation, which is a free parameter in the fit (see posterior distributions in \Fig\ref{fig:deg_scaling}). Since the Faber-Jackson relation is one of the projections of the Fundamental Plane, a few galaxies may lie above this relation, as more compact galaxies tend to have higher velocity dispersions for a given luminosity.

Following B19, by assuming a fixed scaling between the total mass of the cluster galaxies and their luminosity, that is $M_{tot,i}/L_i\propto L_i^{\gamma}$, the slope of the scaling relation in Eq.\ref{eq.: Scaling_relation_rcut} can be determined by:
\begin{equation}
    \beta_{cut}=\gamma-2\alpha+1,
    \label{eq.: slopes}
\end{equation}

\noindent where $\gamma=0.2$ to be consistent with the observed fundamental plane relation.
In \T\ref{table:kinemacs_scaling_relations}, we quote the best-fit values of $\alpha$, $\sigma_{ap}^{ref}$, $\Delta\sigma_{ap}$, and the inferred value of $\beta_{cut}$.

In our lens models, we fix the two slopes $\alpha$ and $\beta_{cut}$ to the fitted (or inferred) values, while we use a Gaussian prior for the normalization $\sigma_{LT}^{ref}$ centered on the measured $\sigma_{ap}^{ref}$, with standard deviation equal to $\Delta\sigma_{ap}$. To obtain this prior on $\sigma_{LT}^{ref}$, we deproject the measured $\sigma_{ap}^{ref}$, as detailed in B19. Conversely, a large uniform prior between 1\arcsec\ and 50\arcsec\ is adopted for the normalization $r_{cut}^{ref}$ in \Eq\ref{eq.: Scaling_relation_rcut}.

In the four lens models described here, subhalos associated to 212 out of 213 cluster galaxies are drawn from the scaling relations, whereas the galaxy \IDFB\ (RA=4:16:08.18, Dec=$-$24:04:00.28) is modeled as an additional elliptical dPIE with negligible core radius. This galaxy, belonging to the Sys-14, produces eight multiple images associated to the families 14.1 and 14.4 (see \Sec\ref{sec:pre_images}). 
In the \BCGs\ model, also the  BCGs are optimized outside the scaling relations as circular core-less dPIE profiles.
Other details on the different mass components included in the four lens models are reported in \T\ref{table:lens_models}.

\begin{figure}[t!]
	\centering
	\includegraphics[width=\linewidth]{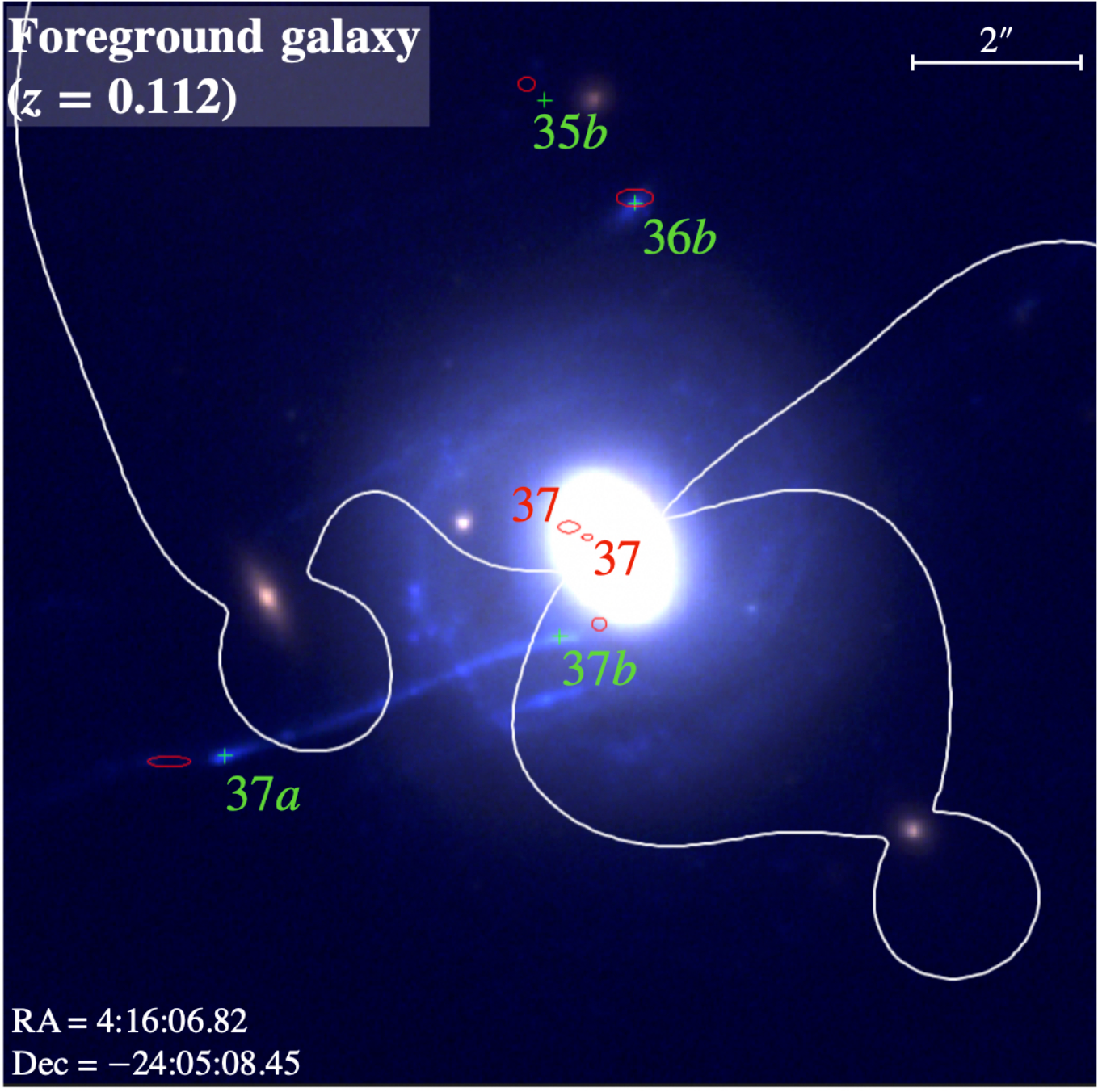}
	\caption{RGB cut-out (F814W, F606W, F435W) centered on the foreground galaxy at $z=0.112$ located $\sim\!30\arcsec$ at SW of the BCG-S (RA=4:16:06.82, Dec=$-$24:05:08.45). The color-scale of the image is adjusted to emphasize the lensed structures behind the galaxy. Green crosses mark the positions of the observed multiple images, while red ellipses show their predicted positions by the \qth\ reference lens model. The sizes of the ellipses refer to $1\mbox{-}\sigma$ errors along the $x$ and $y$ directions. The white line is the critical line computed for a source at $z=2.218$, which is the redshift of family 37.}
	\label{fig:FG_gal}
\end{figure}

In the top panel of \T\ref{table:inout_lensing}, we report the set of input parameters of the \qth\ model, including the range of the flat priors adopted for those which are left to vary. The four dPIE profiles describing the cluster-scale mass distribution introduce 22 free-parameters in the model (see \Sec\ref{sec: clusterscale_mass_distribution}). We also indicate the fixed parameters of the four dPIEs used to model the hot-gas mass distribution. The subhalo mass components includes eight additional free-parameters in this model. Two parameters describe the foreground galaxy at $z=0.112$ (see \Fig\ref{fig:FG_gal}) and two are associated to the normalizations of the scaling relations. 
 The Gaussian prior adopted for $\sigma_{LT}^{ref}$, which is derived  from the measured stellar kinematics of the cluster members (see above), is indicated in square brackets. In conclusion, the \qth\ lens model includes a total of $N_{par}=30$ free parameters, with $N_{con}=232$ constrains corresponding to 202 DoF (see \Eq\ref{eq.: DoF}).

\begin{table*}[h!]  
	\tiny
	\def\arraystretch{1.6}
	\centering    
	\begin{tabular}{|c|c|c|c|c|c|c|>{\centering\arraybackslash}m{9cm}|}
	   \hline
	   \multicolumn{8}{|c|}{\textbf{\normalsize Properties of selected lens models}}\\[2pt] 
	   \hline
	   \textbf{\normalsize Model ID} & \boldmath{\normalsize $N_{par}$} & \textbf{\normalsize DoF} & \boldmath{\normalsize $\Delta_{rms}\,\mathrm{[\arcsec]}$} &
	   \textbf{\normalsize BIC} &
	   \textbf{\normalsize AIC} &
	   {\normalsize $\boldsymbol{\chi^2_{kin}}$} &
	   \textbf{\normalsize Description} \\[2pt] 
	   \hline
	   \textbf{\qth} & \textbf{30} & \textbf{202} & \textbf{0.40} & \textbf{346} & \textbf{243} & \textbf{4941} & \textbf{Four cluster-scale halos, 212 cluster members including BCGs} \cr
	   \hline
	   $\BCGs$ & 28 & 204 & 0.45 & 408 & 311 & 6653 & Three cluster-scale halos, 210 cluster members excluding BCGs \cr
	   \hline
	   \texttt{LM-HLBCGs} & 28 & 204 & 0.46 & 412 & 316 & 7217 & Three cluster-scale halos, 212 cluster members including BCGs \cr
	   \hline
	   \texttt{LM-SHEAR} & 27 & 205 & 0.48 & 413 & 320 & 5042 & Three cluster-scale halos, one shear term, one convergence term, 212 cluster members including BCGs \cr
	   \hline
	\end{tabular}
	\smallskip
    \caption{Description of the four selected best lens models. $N_{par}$ and DoF are the number of model free-parameters and degrees-of-freedom. $\Delta_{rms}$ is the root-mean-square displacement between the positions of observed and model-predicted multiple images (see \Eq\ref{eq.: rms_lt}). The BIC (Bayesian Information Criterion) and AIC (Akaike Information Criterion) values are computed using \Eq\ref{eq: BICAIK}. The $\chi^2_{kin}$ value, defined in \Eq\ref{eq.: kinchi}, quantifies the agreement between the model predicted and measured cluster member velocity dispersions. In the last column, we summarize the differences of the mass parametrization in the four lens models. The reference model selected on the basis of the best figure of merits (from column four to seven) is indicated in bold.
    }
	\label{table:lens_models} 

\end{table*}

\begin{table*}[]     
	\tiny
	\def\arraystretch{2.3}
	\centering          
	\begin{tabular}{|c|c|c|c|c|c|c|c|c|}
	    \cline{3-9}
		\multicolumn{2}{c|}{} & \multicolumn{7}{c|}{ \textbf{Input parameter values and intervals of the \qth\ lens model}} \\
		\cline{3-9}
		  \multicolumn{2}{c|}{} & \boldmath{$x\, \mathrm{[arcsec]}$} & \boldmath{$y\, \mathrm{[arcsec]}$} & \boldmath{$e$} & \boldmath{$\theta\ [^{\circ}]$} & \boldmath{$\sigma_{LT}\, \mathrm{[km\ s^{-1}]}$} & \boldmath{$r_{core}\, \mathrm{[arcsec]}$} & \boldmath{$r_{cut}\, \mathrm{[arcsec]}$} \\ 
          \hline

		  \multirow{8}{*}{\rotatebox[origin=c]{90}{\textbf{Cluster-scale halos}}} 
		  
		  & \boldmath{$1^{st}$} \bf{Cluster Halo} & $\left[-15.0,\,15.0\right]$ & $\left[-15.0,\,15.0\right]$ & $\left[0.2,\,0.9\right]$ & $\left[100.0,\,180.0\right]$ & $\left[350.0,\,1000.0\right]$ & $\left[0.0,\,20.0\right]$ & 2000.0 \\
		  
		  & \boldmath{$2^{nd}$} \bf{Cluster Halo} & $\left[15.0,\,30.0\right]$ & $\left[-45.0,\,-30.0\right]$ & $\left[0.2,\,0.9\right]$ & $\left[90.0,\,170.0\right]$ & $\left[350.0,\,1200.0\right]$ & $\left[0.0,\,25.0\right]$ & 2000.0  \\
		  
		  & \boldmath{$3^{rd}$} \bf{Cluster Halo} & $\left[-55.0,\,-25.0\right]$ & $\left[0.0,\,30.0\right]$ & $0.0$ & $0.0$ & $\left[50.0,\,750.0\right]$ & $\left[0.0,\,35.0\right]$ & 2000.0  \\
		  
		  & \boldmath{$4^{th}$} \bf{Cluster Halo} & $\left[-10.0,\,50.0\right]$ & $\left[-75.0,\,-15.0\right]$ & $\left[0.2,\,0.9\right]$ & $\left[0.0,\,180.0\right]$ & $\left[100.0,\,1000.0\right]$ & $\left[0.0,\,20.0\right]$ & 2000.0 \\
		  \cline{2-9}
		  
		  & \boldmath{$1^{st}$} \bf{Gas Halo} & $-$18.14 & $-$12.13 & 0.12 & $-$156.76 & 433.0 & 149.21 & 149.82  \\[-2ex]
		  
		  & \boldmath{$2^{nd}$} \bf{Gas Halo} & 30.79 & $-$48.67 & 0.42 & $-$71.50 & 249.0 & 34.77 & 165.77  \\[-2ex]
		  
		  & \boldmath{$3^{rd}$} \bf{Gas Halo} & $-$2.37 & $-$1.26 & 0.42 & $-$54.74 & 101.7 & 8.28 & 37.59  \\[-2ex]
		  
		  & \boldmath{$4^{th}$} \bf{Gas Halo} & $-$20.13 & 14.74 & 0.40 & $-$49.32 & 281.8 & 51.67 & 52.34  \\

          \hline
          \multicolumn{1}{c}{}
          \\[-5ex]

          \hline

		  \multirow{3}{*}{\rotatebox[origin=c]{90}{\textbf{Subhalos}}} 
		  
		  & $\boldsymbol{Gal\mbox{-}8971}$ \textbf{(Sys-14)} & $13.35$ & $2.62$ & $\left[0.0,\,0.6\right]$ & $\left[-90.0,\,90.0\right]$ & $\left[60.0,\,200.0\right]$ & $0.0001$ & $\left[0.0,\,50.0\right]$ \\
		  
		  \cline{2-9}
		  
		  & \bf{Foreground gal.} & $31.96$ & $-65.55$ & $0.0$ & $0.0$ & $\left[50.0,\,350.0\right]$ & $0.0001$ & $\left[5.0,\,100.0\right]$ \\
		  
		  \cline{2-9}
		  
		  & \bf{Scaling relations} & $\boldsymbol{N_{gal}=}212$
		  & $\boldsymbol{m_{F160W}^{ref}=}17.02$
		  & $\boldsymbol{\alpha=}0.30$ & $\boldsymbol{\sigma_{LT}^{ref}=}\left(248\pm28\right)$ & $\boldsymbol{\beta_{cut}=}0.60$ & $\boldsymbol{r_{cut}^{ref}=}\left[1.0,\,50.0\right]$ & $\boldsymbol{\gamma=}0.20$\\
	
          \hline
          
	\end{tabular}
	\\[6ex]
	\begin{tabular}{|c|c|c|c|c|c|c|c|c|}
	    \cline{3-9}
		\multicolumn{2}{c|}{} & \multicolumn{7}{c|}{ \textbf{Optimized parameters of the \qth\ lens model}} \\
		\cline{3-9}
		  \multicolumn{2}{c|}{} & \boldmath{$x\, \mathrm{[arcsec]}$} & \boldmath{$y\, \mathrm{[arcsec]}$} & \boldmath{$e$} & \boldmath{$\theta\ [^{\circ}]$} & \boldmath{$\sigma_{LT}\, \mathrm{[km\ s^{-1}]}$} & \boldmath{$r_{core}\, \mathrm{[arcsec]}$} & \boldmath{$r_{cut}\, \mathrm{[arcsec]}$} \\ 
          \hline

		  \multirow{4}{*}{\rotatebox[origin=c]{90}{\textbf{Cluster-scale halos}}} 
		  
		  & \boldmath{$1^{st}$} \bf{Cluster Halo} & $-0.42_{-0.53}^{+0.58}$ & $-0.14_{-0.39}^{+0.37}$ & $0.85_{-0.01}^{+0.01}$ & $144.5_{-0.6}^{+0.5}$ & $560.7_{-18.2}^{+20.5}$ & $7.6_{-0.6}^{+0.6}$ & 2000.0 \\
		  
		  & \boldmath{$2^{nd}$} \bf{Cluster Halo} & $25.41_{-1.28}^{+0.99}$ & $-38.21_{-1.28}^{+1.11}$ & $0.87_{-0.04}^{+0.02}$ & $132.0_{-1.2}^{+1.4}$ & $645.8_{-40.6}^{+76.4}$ & $12.5_{-0.9}^{+1.1}$ & 2000.0  \\
		  
		  & \boldmath{$3^{rd}$} \bf{Cluster Halo} & $-35.32_{-1.77}^{+1.51}$ & $10.02_{-1.07}^{+1.10}$ & 0.0 & 0.0 & $384.7_{-26.5}^{+22.4}$ & $12.1_{-1.6}^{+1.2}$ & 2000.0  \\
		  
		  & \boldmath{$4^{th}$} \bf{Cluster Halo} & $18.84_{-1.05}^{+0.66}$ & $-44.58_{-1.14}^{+1.13}$ & $0.78_{-0.03}^{+0.05}$ & $117.0_{-4.9}^{+2.4}$ & $562.8_{-105.2}^{+40.1}$ & $10.9_{-0.9}^{+0.9}$ & 2000.0 \\
		  \cline{2-9}
		  
          \hline
          \multicolumn{1}{c}{}
          \\[-5ex]
          
          \hline
		  
		  \multirow{3}{*}{\rotatebox[origin=c]{90}{\textbf{Subhalos}}} 
		  
		  & $\boldsymbol{Gal\mbox{-}8971}$ \textbf{(Sys-14)} & 13.35 & 2.62 & $0.51_{-0.10}^{+0.06}$ & $-40.8_{-9.6}^{+12.1}$ & $111.4_{-4.5}^{+4.1}$ & 0.0001 & $29.8_{-12.6}^{+10.8}$ \\
		  
		  \cline{2-9}
		  
		  & \bf{Foreground. gal.} & 31.96 & $-65.55$ & 0.0 & 0.0 & $138.7_{-12.7}^{+19.6}$ & 0.0001 & $49.6_{-20.1}^{+28.4}$ \\
		  
		  \cline{2-9}
		  
		  & \bf{Scaling relations} & $\boldsymbol{N_{gal}=}212$
		  & $\boldsymbol{m_{F160W}^{ref}=}17.02$
		  & $\boldsymbol{\alpha=}0.30$ & $\boldsymbol{\sigma_{LT}^{ref}=}252.0_{-5.0}^{+5.5}$ & $\boldsymbol{\beta_{cut}=}0.60$ & $\boldsymbol{r_{cut}^{ref}=}11.2_{-1.3}^{+1.5}$ & $\boldsymbol{\gamma=}0.20$\\
	
          \hline
          
	\end{tabular}
	
	\smallskip
    \caption{
    {\it Top:} Input parameters of the \qth\ reference lens model. Singles numbers refer to fixed parameters. For the free parameters, we quote within square brackets the boundaries of the input flat priors. For the  $\sigma_{LT}^{ref}$ parameter, a Gaussian prior is adopted with mean and standard deviation indicated in round brackets. $N_{gal}$ is the number of cluster member galaxies included in the scaling relations (see Eqs.\,\ref{eq.: Scaling_relation_sigma} and \ref{eq.: Scaling_relation_rcut}). 
    {\it Bottom:} Optimized output values of the free parameters of the reference lens model. We quote the median value and the [16-th, 84-th] percentiles from the marginalized posterior distribution.
    }    

	\label{table:inout_lensing}

\end{table*}

\begin{figure}[ht!]
	\centering
	\includegraphics[width=\linewidth]{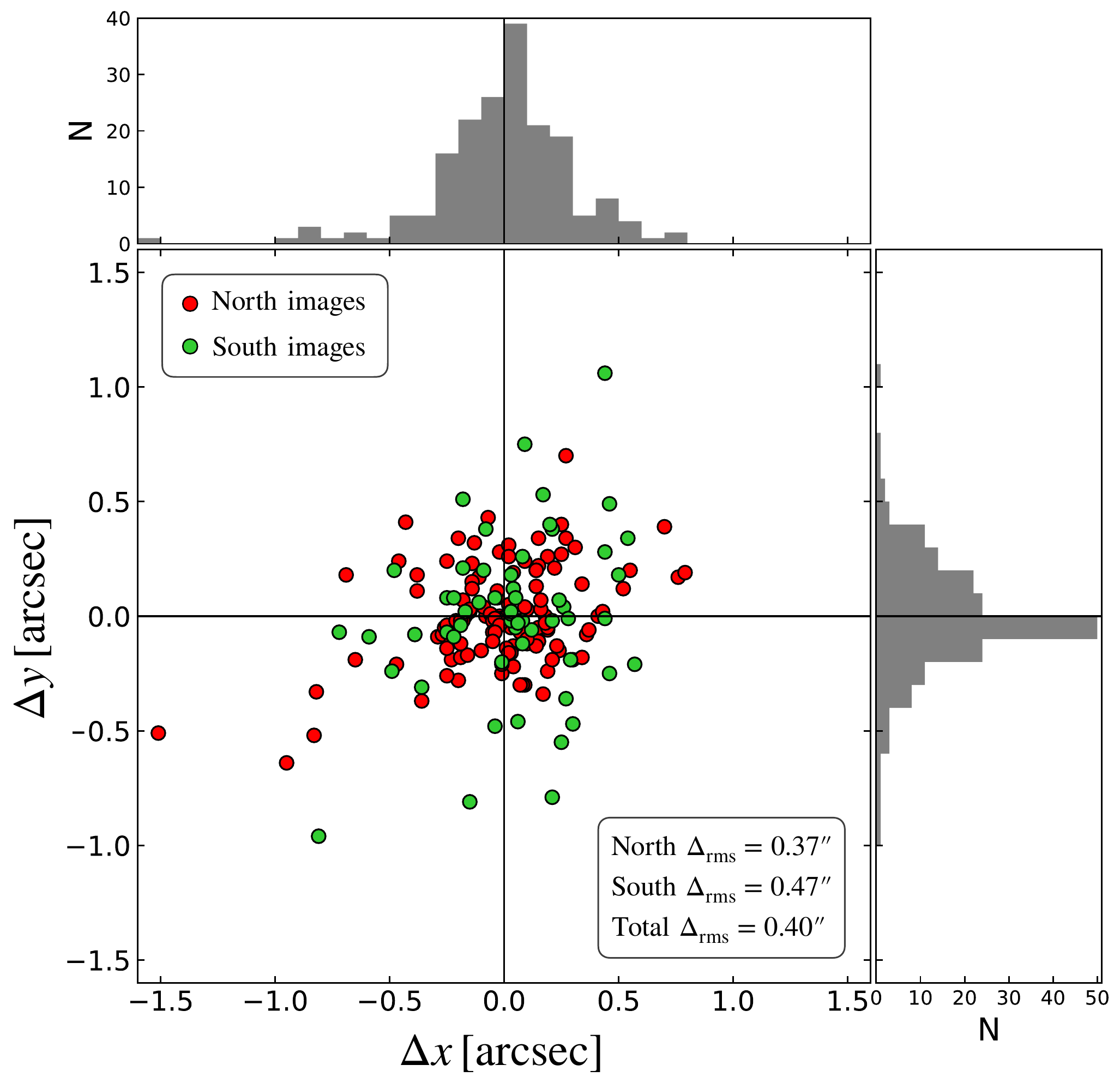}	\caption{2D and 1D distributions of the displacements $\Delta_i$, along $x$ and $y$ directions, between the observed and model-predicted positions of the 182 multiple images used to constrain the \qth\ lens model, and corresponding $\Delta_{rms}$ values (see \Eq\ref{eq.: rms_lt}). The 125 images in the north field and the 57 in the south are shown separately as red and green dots, respectively (see the dividing line between the NE and SW regions in \Fig\ref{fig:DM_images}).}
	\label{fig:rms}
\end{figure}

\section{Results}
\label{sec:res}
The main results from the optimizations of the four lens models are summarized in \T\ref{table:lens_models}. In addition to the  three indicators of the goodness of the lens models defined above ($\Delta_{rms}$, BIC, and AIC), we introduce a fourth figure of merit, $\chi^{2}_{kin}$, which quantifies the agreement between the lens predicted and measured stellar velocity dispersions of member galaxies. If we call $\sigma_{ap}^{SR}(m_{F160W,i})$ the aperture-averaged projected velocity dispersion of the $i$-th galaxy (with luminosity $m_{F160W,i}$), inferred from the best-fit \LT\ scaling relation, $\chi^{2}_{kin}$ is defined as:

\begin{equation}
    \chi^{2}_{kin}=\sum_{i=1}^{N_{m}^{gal}} \left(\frac{\sigma_{ap}^{SR}(m_{F160W,i}^{gal})-\sigma_{ap,i}^{gal}}{d\sigma_{ap,i}^{gal}}\right)^2 ,
    \label{eq.: kinchi}
\end{equation}
 
\noindent where $N_{m}^{gal}$ is the number of galaxies with a measured velocity dispersion $\sigma_{ap}^{gal}$. As general rule, the lower is the $\chi^{2}_{kin}$ value, the better is the agreement between the \LT\ best-fit $\sigma\mbox{-}mag$ scaling relation and the measured Faber-Jackson relation.

Based on these four figure of merits, the \qth\ lens model emerges as the best model, which reproduces the positions of the multiple images with the lowest $\Delta_{rms}$, and best match the internal kinematics of the cluster member galaxies (lowest $\chi^{2}_{kin}$). This model also possesses the lowest values of the BIC and AIC criteria. In the upcoming sections, we therefore characterize in detail this reference model by discussing its ability to predict robust positions and magnifications of multiple images, specifically those close to critical lines and around selected cluster galaxies. 

\subsection{\qth\ lens model: predicted positions and magnifications of the multiple images}
\label{sec:pre_images}

\begin{figure}[t!]
	\centering
	\includegraphics[width=\linewidth]{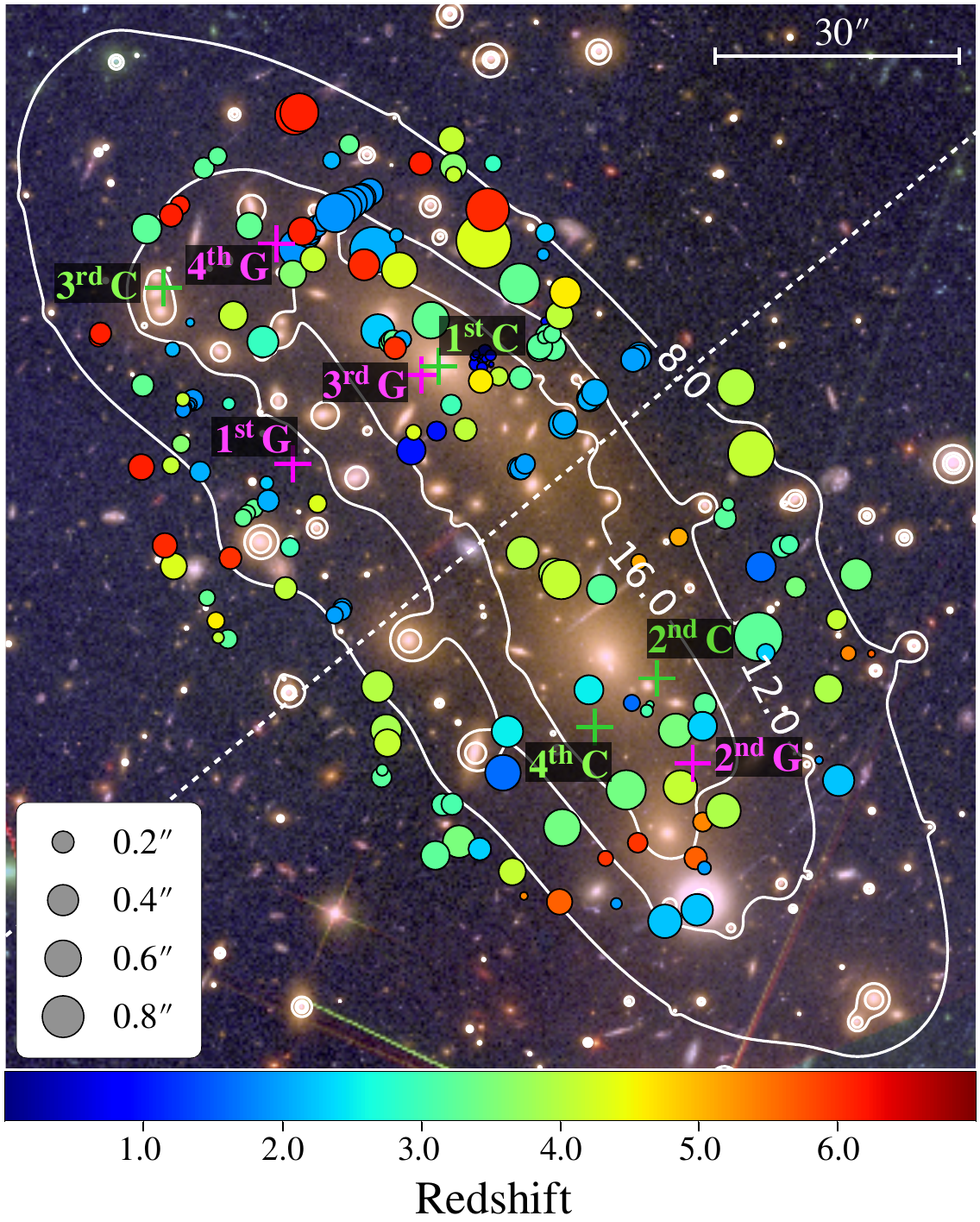}
	\caption{RGB image of \CL\ (as in \Fig\ref{fig:CM}), with overlaid contours (in white) of the total projected mass distribution obtained from the reference lens model. The contours are expressed in units of $\mathrm{10^{8}\,M_{\odot}\,kpc^{-2}}$. The magenta crosses mark the positions of the four dPIE profiles used to describe the hot-gas component of the cluster. The green crosses mark the centers of the four cluster-scale halos. The circles show the positions of the 182 observed multiple images, with colors encoding their redshift. Their sizes scale proportionally to the displacements between the observed and model-predicted positions of the multiple images ($\left\|\boldsymbol{\Delta_i} \right\|$). The white dashed line separates two sets of multiple images in the northern and southern fields.}
	\label{fig:DM_images}
\end{figure}

In \Fig\ref{fig:rms}, we show the distribution of the differences between the observed and model-predicted positions of the multiple images in the $x$ and $y$ directions and corresponding values of $\Delta_{rms}$. It is worth noting that the resulting $\Delta_{rms}=0.40\arcsec$ is 34\% smaller than our previous model (B19) despite the significantly larger number of multiple images (182 vs 102). 
The $\Delta_{rms}$ value in the MDLF field (0.37\arcsec) is appreciably smaller than the one in the southern field (0.47\arcsec). We interpret this as the result of a better constrained mass model in the NE field due to the large number of multiple images \citep[see][]{Vanzella_2020}.  The mass distribution in the SE region appears more complex, as discussed by \cite{Balestra_2016} who found \CL\ in a pre-merging phase based on a dynamical and structural analysis of the cluster. 
Interestingly, we find that the inclusion of the fourth cluster-scale halo in the \qth\ lens model is needed to significantly reduce $\Delta_{rms}$ in the SE field, thus confirming the complex structure of the southern cluster field. The bright foreground galaxy (\Fig\ref{fig:FG_gal}), which we simply model as a circular dPIE at the cluster redshift, adds further uncertainty to the projected mass distribution in this region. In \Fig\ref{fig:DM_images}, we show the spatial distribution of the absolute displacements, $\Delta_i$, between the observed and model-predicted positions of multiple images. This also shows that the multiple images in the NE field are, on average, better reproduced independently from their redshifts.

The reference lens model predicts 226 multiple images associated to 66 independent sources. 
We note that there are 44 additional images predicted ``a posteriori'' by the model, which are still waiting for a secure identification (see for example Sys-12 at the end of this subsection). 

\begin{figure}[t!]
	\centering
	\includegraphics[width=\linewidth]{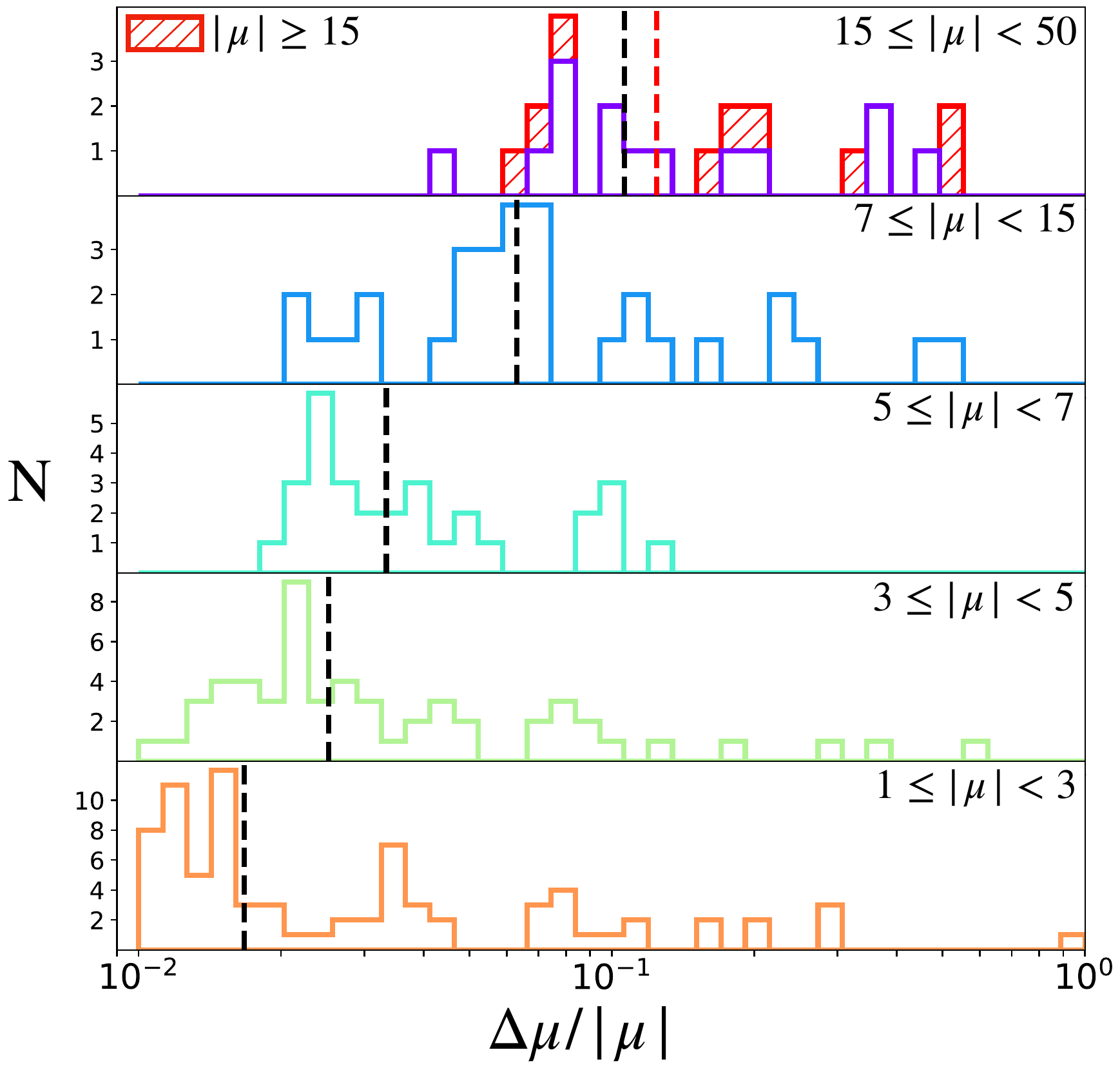}
	\caption{
	Distributions of the relative statistical error on the absolute magnification ($\Delta \mu / \left | \mu \right |$) for the 226 multiple images that are predicted by the reference lens model. Histograms refer to six different intervals of $\left | \mu \right |$.  Vertical black dashed lines mark the median values of each distribution (the red dash line corresponds to bin with $\left | \mu \right | \ge 15 $).
	 }
	\label{fig:magnification_h}
\end{figure}

To study the statistical uncertainty ($\Delta \mu$) on the absolute magnification, $\left | \mu \right |$, we show in \Fig\ref{fig:magnification_h} the relative error on the absolute magnification in different $\left | \mu \right| $ regimes. This is particularly relevant for the study of high-$z$ strongly magnified sources \citep[see][]{Vanzella_2020}. 
The magnification uncertainties are computed as follows.  Firstly, we generate 500 realizations of the \qth\ lens model by randomly extracting 500 parameter samples from the \LT\ MCMC chains. 
For each realization, we compute the absolute magnifications at the predicted position of the multiple images and derive their posterior distributions. To confidently assign the correct magnification to each image, we verify that each model realization predicts the expected parity for that image. We define $\left | \mu \right |$ as the median value of the magnification distributions, and $\Delta \mu$ as half of the difference between the 84-th and 16-th percentiles.

One can see in \Fig\ref{fig:magnification_h} that the relative magnification uncertainty progressively increases at larger $\left | \mu \right |$. 
This is due to the fact that the most magnified images are located closer to the critical lines of the lens model where magnification gradients becomes particularly large. In these regions we also expect that systematic errors due to model parametrization become increasingly important. This analysis is deferred to a future paper.  

\begin{figure}[t!]
	\centering
	\includegraphics[width=\linewidth]{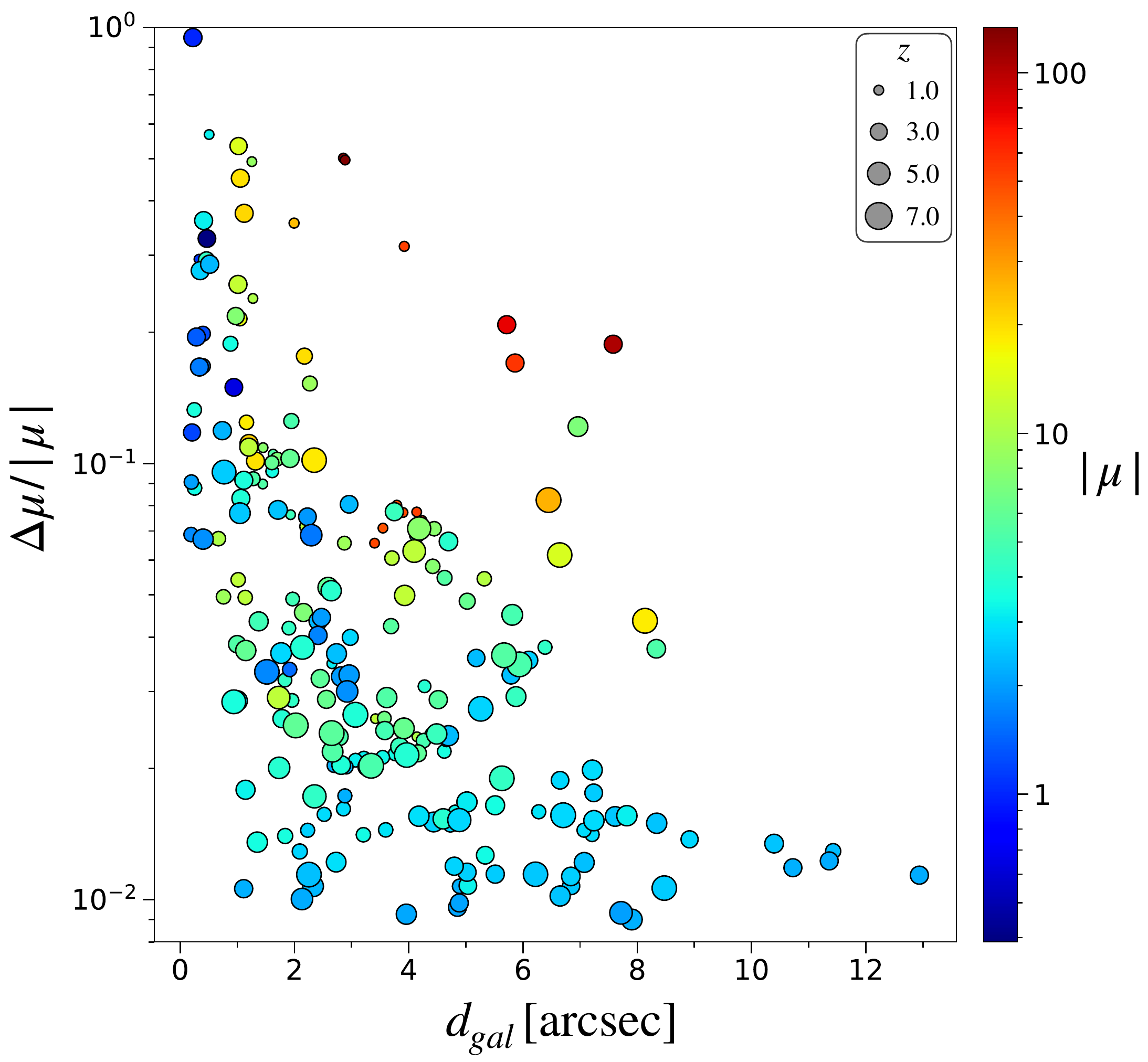}
	\caption{
	Relative statistical uncertainty on the absolute magnification $\Delta \mu / \left | \mu \right |$ for the 226 predicted multiple images (as in \Fig\ref{fig:magnification_h}) as a function of their distances from the closest cluster galaxy. The colors of the circles encode the $\left | \mu \right |$ values, while their sizes scale according to their redshifts.
	 }
	\label{fig:magnification}
\end{figure}

To better understand the origin of magnification uncertainties, we investigate how $\Delta \mu / \left | \mu \right |$ varies as a function of the distance between the multiple images and their closest cluster galaxy, and the magnification itself.  The result is shown in \Fig\ref{fig:magnification}. 
On average, the images that form close to cluster members have larger relative errors. This anti-correlation is explained by the fact that galaxy masses act as small gravitational lenses embedded into the cluster potential and introduce numerous secondary critical lines (with sizes of few arcseconds) into the lens model. Images closer to critical lines tend to have larger $\Delta \mu / \left | \mu \right |$ values. The few highly magnified images with $\Delta \mu / \left | \mu \right |\sim 20\%$ at distances of 6-8\arcsec\ from cluster galaxies are necessarily located around the main critical lines of the cluster computed for their redshifts.

We now define a new metric, which is sensitive to the gradients of the deflection field. 
We focus on a number of specific lensed systems to test the ability of the lens model to predict the positions of multiply imaged clumps, associated to the same resolved source, which are close to the critical lines. 

We take Sys-5 (see bottom-right panels of \Fig\ref{fig:cluster}) as an example. This system is composed by six clumps (5.1, 5.2, 5.3, 5.4, 5.5, and 5.6) belonging to the same extended source at $z=1.893$. Each clump is imaged three times, so that the source 5.1, for example, produces the three images $5.1a$, $5.1b$, and $5.1c$. 
Using the observed and model-predicted positions of the multiple images, we determine the ``distance vectors'' connecting each pair of lensed clumps belonging to the same extended image: for example, $\Vec{d_{obs(pre)}^{5.1a,5.2a}}$, $\Vec{d_{obs(pre)}^{5.1a,5.3a}}$, $\Vec{d_{obs(pre)}^{5.2a,5.3a}}$, etc. (similar combinations are obtained for images $b$ and $c$). The modulus of these vectors corresponds to the distance between these knots. Moreover, we measure the angular difference in the orientation of the corresponding observed and predicted distance vectors, so that $\left|\Delta\phi^{i,j}\right|=\arccos\left[\Vec{\hat{d}_{pre}^{i,j}}\cdot\Vec{\hat{d}_{obs}^{i,j}}\right]$, where we use ``$\cdot$'' to indicate the scalar product, while $i$ and $j$ are different knots (e.g., $\left|\Delta\phi^{5.1a,5.2a}\right|=\arccos\left[\Vec{\hat{d}_{pre}^{5.1a,5.2a}}\cdot\Vec{\hat{d}_{obs}^{5.1a,5.2a}}\right]$).

\begin{figure}[t!]
	\centering
	\includegraphics[width=\linewidth]{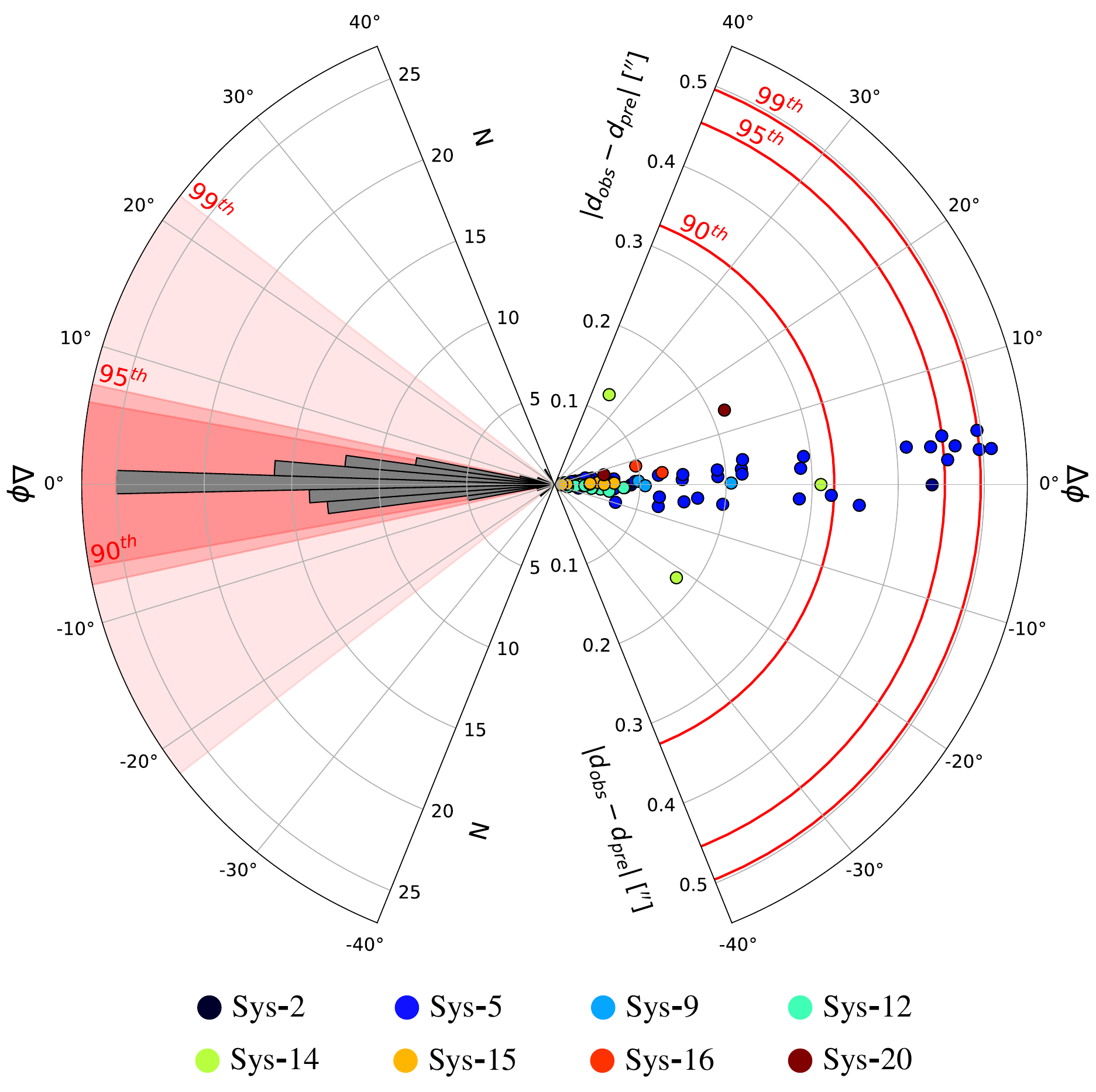}
	\caption{Difference between the observed and model-predicted separations and orientations of each pair of clumps (or knots) identified within eight systems of multiple images (see \Fig\ref{fig:cluster}). Each dot corresponds to a different pair of clumps, colored according to the parent system. We define $\boldsymbol{d_{obs}^{ij}}$ ($\boldsymbol{d_{pre}^{ij}}$) as the measured (predicted) distance between the $i$-th and $j$-th clump belonging to the same resolved lensed image (e.g., $i=12.1c$, $j=12.5c$ in the top left panel of \Fig\ref{fig:SI}). We plot the absolute difference $|\boldsymbol{d_{obs}^{ij}}-\boldsymbol{d_{pre}^{ij}}|$ along the radial axis on the right side, while the angle ($\Delta \phi_{ij}$) between $\boldsymbol{d_{obs}^{ij}}$ and $\boldsymbol{d_{pre}^{ij}}$ vectors (measured counterclockwise) is plotted across the angular coordinate. On the left side of the diagram, we draw a binned distribution of all the data-points. The 90-th, 95-th, and 99-th percentiles of the distributions along the radial and angular directions are shown as red semi-circles and sectors respectively.
	 }
	\label{fig:knots}
\end{figure}

On the right side of \Fig\ref{fig:knots}, we plot, on the radial axis, the absolute value of the difference between the predicted and observed distance vectors ($|\Vec{d_{obs}^{i,j}}-\Vec{d_{pre}^{i,j}}|$) for the whole set of multiple images belonging to eight different systems. The values of $\Delta \phi^{i,j}$ are plotted along the angular coordinate. In this representation, a perfectly reconstructed pair of images should lie on the origin of the reference frame. On the left side of the diagram, we show the angular distribution of the $\Delta\phi^{i,j}$ values. In 90\% of the cases, the difference in the distances between the observed and predicted knots is lower than $0.33\arcsec$. Similarly, the $\Delta \phi^{i,j}$ values are lower than $5.9^{\circ}$ for 90\% of image pairs. This result demonstrates that our reference lens model not only accurately predicts the observed position of the multiple images ($\Delta_{rms}=0.40\arcsec$), but it is also able to accurately reproduce the extension and the orientation of the inner structure of extended multiple images.

\begin{figure}[t!]
	\centering
	\includegraphics[width=\linewidth]{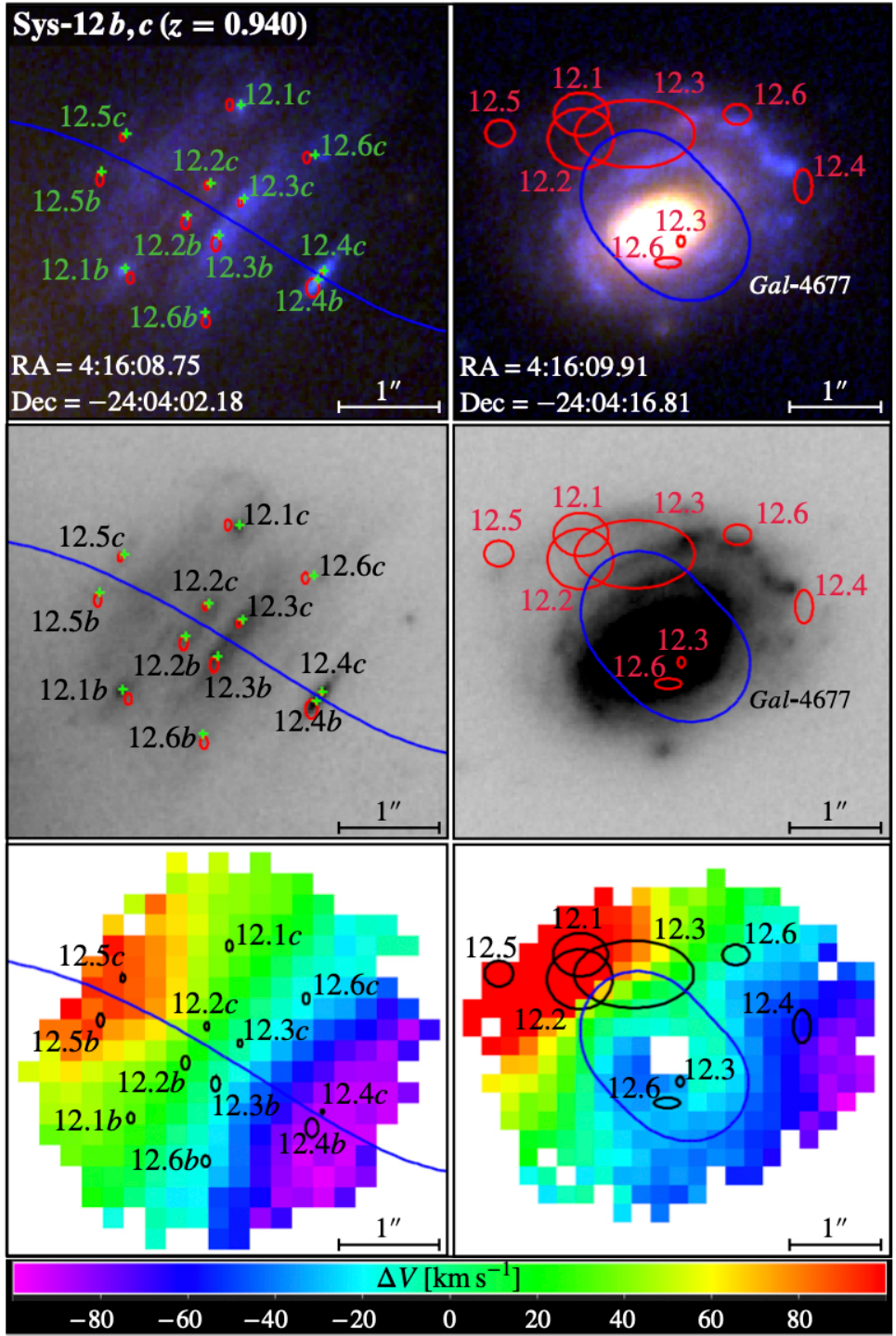}
	\caption{Detailed analysis of the lensed system 12 at $z=0.940$. In the top and middle left panels, six point-like clumps ($12.1\ldots 12.6$) are identified within the lensed images $a$ and $b$, straddling the critical line at $z=0.940$ (blue line). Green crosses mark the positions of the observed multiple images. Red ellipses show the positions predicted by the \qth\ reference lens model, with sizes corresponding to $1\mbox{-}\sigma$ errors along the $x$ and $y$ directions.   The top and middle right panels show the third lensed image predicted by the lens model (blue spiral structure around the cluster galaxy {\it Gal}-4677). The blue line is the associated critical line; the red ellipses correspond to the predicted multiple images. In the top panels, RGB cut-outs combine F814W, 606W, F435W filters, while in the middle panels we show median stack images with the same filters. The bottom panels show the corresponding velocity maps obtained by tracing the shift of the [OII] emission doublet in the MUSE datacube (around $\sim\! 7229\AA$ at $z=0.940$).
	The same red ellipses as in the upper panels are plotted in black for clarity.}
	\label{fig:SI}
\end{figure}

To conclude this session characterizing the robustness of our reference model, we briefly describe the properties of two interesting systems of multiple images, namely Sys-12 and Sys-14.
The extended background source associated to Sys-12, at $z=0.94$, is a spiral galaxy which is lensed into three images by \CL. The top-left and middle-left panels of \Fig\ref{fig:SI} show how two of these images ($b$ and $c$) are resolved into several star-forming clumps which are close to merging onto the critical line. 
\cite{Vanzella_2020} characterized  these systems as star-forming complexes with extremely low luminosity  (M$_{\rm UV} \sim -11$) and small sizes ($\lesssim 30$ pc), thanks to their strong magnification (see below).
We securely identify six point-like knots ($12.1\ldots 12.6$) on each side of the critical line, which are included in our lens model. The clumps with the largest separations, namely 12.4 and 12.5, are 10.9 kpc apart on the source plane, while the closest central knots, 12.2 and 12.3, are only 1.7 kpc apart. The mirrored images 12.4b and 12.4c are located $0.05\arcsec$ from the critical line and have in fact the highest magnifications among the entire sample of multiple images, $\mu_{12.4b}=133_{-46}^{+87}$ and $\mu_{12.4c}=134_{-46}^{+87}$. 
We emphasize how our reference lens model reproduces the 12 positions  of Sys-12 with very high precision ($\Delta_{rms} =0.05\arcsec$), as well as the relative distances and orientations of all knots belonging to a given lensed image ($b$ or $c$), as one can appreciate from \Fig\ref{fig:knots} (cyan dots cluster near the origin of the diagram). 

The third image of Sys-12 is blindly predicted by the lens model $\sim\! 20\arcsec$ SE of image $b/c$ (see \Fig\ref{fig:cluster}), around a cluster member which contributes to create an Einstein ring configuration (see top-right and middle-right panels of \Fig\ref{fig:SI}). This third image is not included in the catalog of multiple images as the corresponding lensed knots cannot be readily identified in HST images. 

The MUSE spectra of the multiple images in Sys-12 are characterized by a prominent [OII]$\lambda$3726.2, 3729.1 emission lines doublet \citep[see][]{Vanzella_2020}. The  wavelength drift of the [OII] peaks across the lensed images can be readily detected in the MUSE datacube\footnote{We use the MUSE Python data analysis framework (MPDAF) for this analysis (https://mpdaf.readthedocs.io/en/latest).}. This analysis yields the velocity maps shown in the lower panels of \Fig\ref{fig:SI}.
By ray-tracing the six star-forming knots in image $b$ and $c$ on the source plane at $z=0.94$, we find them very well aligned with a symmetric rotation curve stretching from $\sim -100\,\mathrm{km\,s^{-1}}$ for 12.4, up to $\sim 100\,\mathrm{km\,s^{-1}}$ for 12.5, across $\sim 11$ kpc. 
We defer to a future publication a detailed study of this and other galaxy-scale systems of multiple images. 

\begin{figure}[t!]
	\centering
	\includegraphics[width=\linewidth]{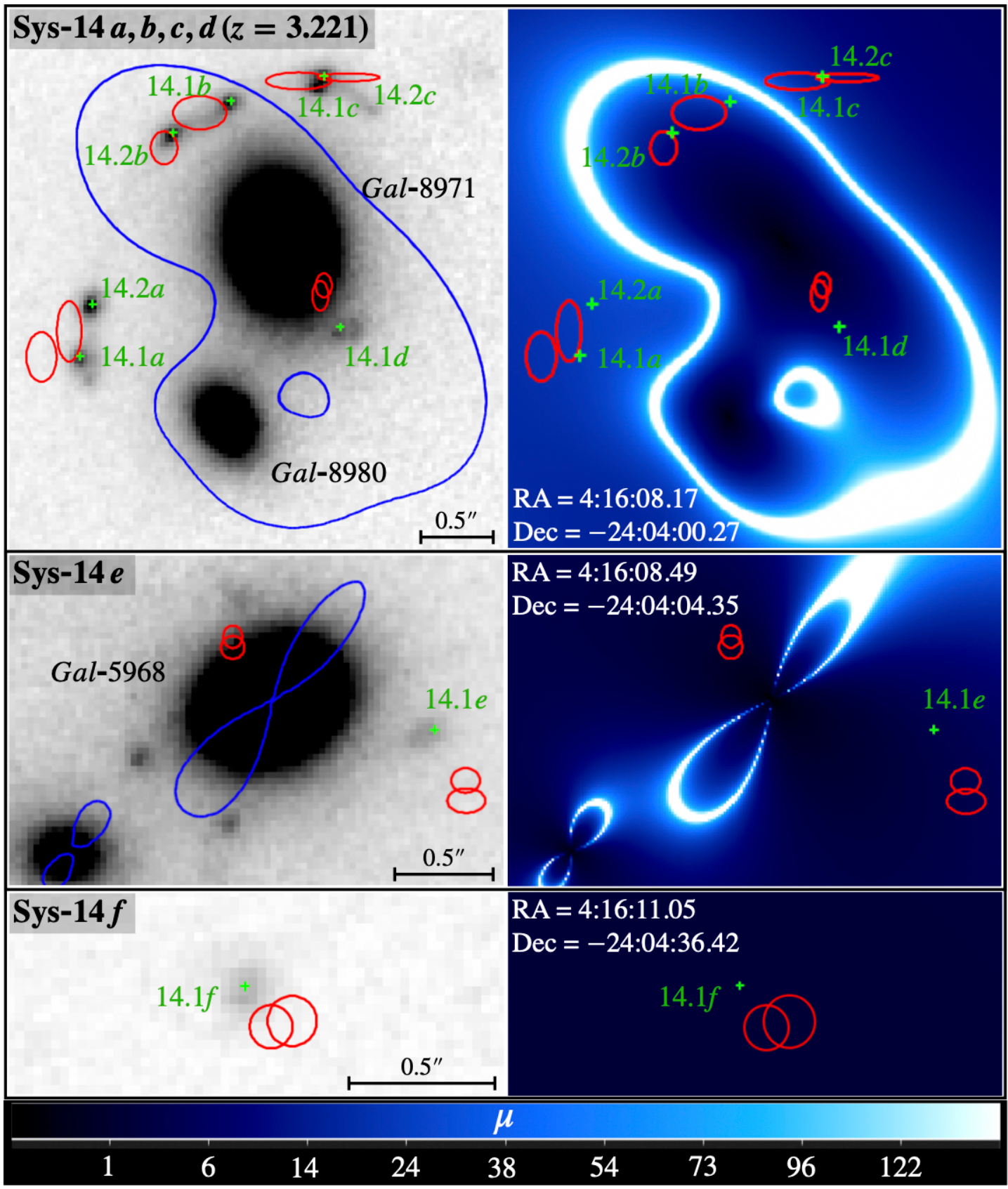}
	\caption{Inspection of system 14  at $z=3.221$. A pair of nearly point-like clumps ($14.1$ and $14.2$) is imaged six times by the cluster mass distribution, around two cluster galaxies ($Gal\mbox{-}8971$ and $Gal\mbox{-}8980$, with F160W magnitudes 19.6 and 21.5, respectively). Green crosses, red ellipses, and blue lines have the same meaning as \Fig\ref{fig:SI}. The three gray-scale cut-outs on the left are median stack images of F814W, F606W, and F435W HST filters. Right panels show  absolute magnification ($|\mu|$) maps computed at $z=3.221$. The lens model predicts a fifth ($e$) and a sixth ($f$) image (middle and bottom panels, see also \Fig\ref{fig:cluster}).}
	\label{fig:ID14}
\end{figure}

Sys-14 (see \Fig\ref{fig:ID14}) was studied in detail by \cite{Vanzella_ID14}. In this system, two close compact sources (14.1 and 14.2) at $z=3.221$ are lensed into six multiple images each (nine out of twelve multiple images are used as constraints into the lens models). Eight images, $14.1a,b,c,d$ and $14.2a,b,c,d$, are located around a pair of cluster galaxies.
Since the mass distribution of the brighter galaxy ($Gal\mbox{-}8971$) strongly affects the geometry of the GGSL, its mass halo is parameterized as an elliptical dPIE profile outside the subhalo scaling relations in all lens models (see \Sec\ref{sec:mass_comp} and \T\ref{table:inout_lensing}). We note that the three images $14.2d$, $14.2e$, and $14.2f$, are not used in the lens models, due to the lack of a secure counterpart in the HST images. Instead, we tentatively identify the $14.2c$ image and use it in the lens model with the same position as $14.1c$ and a positional error which includes both images.
Compared to our previous lens model by B19, the new model predicts an additional counter-image of the 14.1 and 14.2 pair, which however still remains not securely identified (mid panel of \Fig\ref{fig:ID14}). For all images associated to Sys-14, we find $\Delta_{rms} =0.25\arcsec$. We refer to \cite{Vanzella_2020} for a discussion on the magnification and physical sizes of the multiple images of Sys-14. 

\subsection{\CL\ total projected mass distribution}
\label{sec:mass_dist}
As described in \Sec\ref{sec:mass_comp}, the mass distribution of  \CL\ obtained with the \qth\ lens model includes a cluster-scale component and a subhalo population traced by the cluster members. The former includes a fixed hot-gas component and the dominant DM halo which is modeled with four (elliptical) dPIE profiles. The latter includes 212 circular dPIEs with vanishing core radii and a separate elliptical subhalo to model the bright member in Sys-14 (\Fig\ref{fig:ID14}). In \T\ref{table:inout_lensing}, we quote all the input parameters and the output optimized values of each parameter obtained with of our reference  model. In \Fig\ref{fig:DM_images}, we show the iso-density contours of the total projected mass distribution overlaid on the HST image of \CL, and we also indicate the centers of the cluster-scale halos. 

As in C17 and B19, the new mass model does not imply any significant offset between the position of the cluster-scale halo surrounding the northern BCG and the peak of the light distribution (the projected distance of the halo mass peak from the BCG-N is formally $0\arcsec{.}8_{-0.3}^{+0.4}$).
As discussed above, the mass distribution around the southern clump of \CL\ is not sufficiently well constrained to investigate possible offsets with the luminous mass. 
The optimized position of the third halo in the NE region, whose presence is needed to reduce significantly the $\Delta_{rms}$ value, is 
very close to a local over-density of the projected galaxy distribution (\Fig\ref{fig:DM_images}), in keeping with C17 and B19. 

\begin{figure}[t!]
	\centering
	\includegraphics[width=\linewidth]{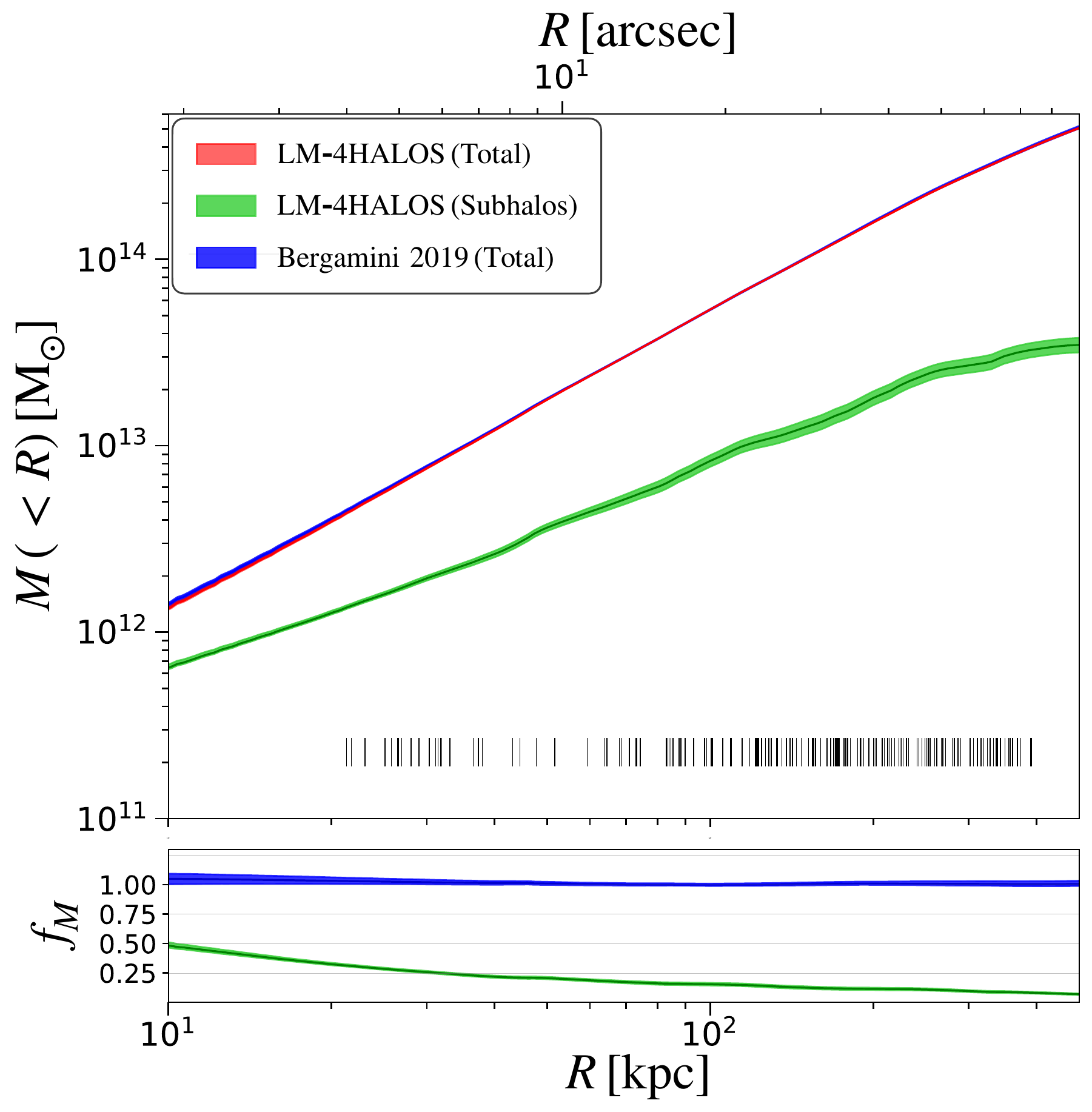}
	\caption{{\it Top:} Cumulative projected total mass profile of \CL\ as a function of the distance ($R$) from the northern BCG. The red curve refers to the median and the $68\%$ confidence level obtained from the reference lens model. The previous result by B19 is plotted in blue. The total mass of the subhalos, associated to cluster galaxies, is shown in green. Black vertical segments mark the positions of the 182 multiple images. {\it Bottom:} Ratio between the new mass profile and the one from B19 (blue) and fractional contribution of the subhalo component to the total cumulative mass (in green).
	 }
	\label{fig:mass}
\end{figure}

The cumulative projected total mass profile of \CL\ as a function of the projected distance, $R$, from the BCG-N is shown in \Fig\ref{fig:mass} (red thin band). The small $1\mbox{-}\sigma$ uncertainties are computed by randomly extracting parameters from the MCMC chains of 500 realizations of the \qth\ model. 
The new cumulative mass profile is found in very good agreement with that obtained with our previous B19 model, which had a somewhat different parametrization and used 80 fewer multiple images (the difference is less than 3\% over the radial range  where the multiple images are located). 

\Fig\ref{fig:mass} also shows the cumulative projected mass profile associated to the subhalos, and their relative contribution to the total mass. At small projected distances from the BCG-N, the subhalos contribute for more than $40\%$ to the total mass, while this contribution does not exceed 15\% at $R> 100$ kpc.

As discussed in \Sec\ref{sec: galaxy_mass_distribution}, the normalization of the $\sigma\mbox{-}m_{F160W}$ scaling relation for the subhalos in all the lens models is obtained adopting a Gaussian prior derived from the observed  $\sigma_{ap}\mbox{-}m_{F160W}$ Faber-Jackson relation.

The resulting scaling relation obtained from the optimization of the \qth\ lens model is shown in \Fig\ref{fig:green_plots} (see red band). The very good agreement between the inferred scaling relation and the observed one (data points and green band) is not common to the other models we analyzed. Indeed, the normalization $\sigma_{LT}^{ref}$ of these secondary models tend to lie significantly below the observed value despite the adopted prior, as it can be appreciated from their higher $\chi^2_{kin}$ values (see \T\ref{table:lens_models} and \Eq\ref{eq.: kinchi}).

In the effort to accurately reproduce the positions of the multiple images associated to the galaxy scale system 14 (see \Fig\ref{fig:ID14}), we have parameterized the mass of \IDFB\ as an extra elliptical dPIE outside the scaling relations (see \Sec\ref{sec: galaxy_mass_distribution}). Its measured velocity dispersion, $169.8\pm 3.5\,\mathrm{km\,s^{-1}}$ (see  magenta triangle in \Fig\ref{fig:green_plots}), is in tension with the aperture-projected  velocity dispersion obtained from the posterior distributions of the \qth\ model ($133.0_{-5.2}^{+5.5}\,\mathrm{km\,s^{-1}}$, see magenta square). Despite the high-quality data of this study, the complex geometry of the mass distribution of this system, with a close galaxy companion embedded in the cluster halo, makes it difficult to reliably infer both the values of the velocity dispersion and truncation radius of $Gal\mathrm{-}8971$ from the lens model. 

The projected mass of a dPIE within an aperture of radius $R_E$ is given by (\citealt{Eliasdottir_lenstool} and B19):

\begin{equation}
\label{Eq.:mass_dPIE}
    M(R_E)=\frac {\pi \sigma_0^2}{G}\left(\sqrt{r_{core}^2+R_E^2}-r_{core}-\sqrt{r_{cut}^2+R_E^2}+r_{cut}\right).
\end{equation}

\noindent For a fixed value of the core radius, a higher values of $r_{cut}$ produces the same  mass $M(R_E)$ for a smaller central velocity dispersion $\sigma_0$. Our best fit model yields an $r_{cut}$ value for this specific galaxy exceeding $R_E$, with a large uncertainty.

\begin{figure}[t!]
	\centering
	\includegraphics[width=\linewidth]{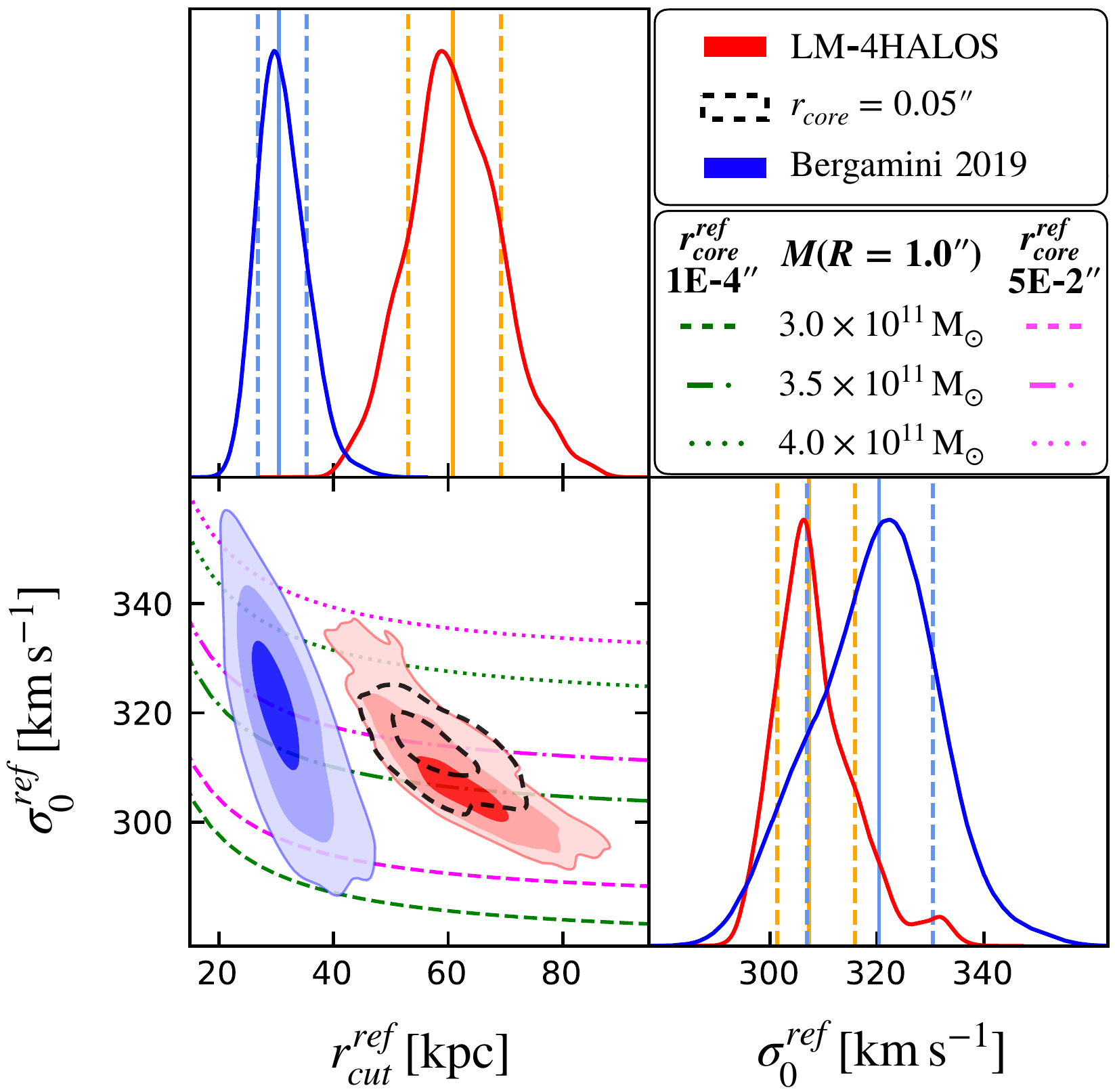}
	\caption{Marginalized posterior distributions of the normalizations $\sigma_0^{ref}$ and  $r_{cut}^{ref}$ of the cluster member scaling relations (see \Eq\ref{eq.: Scaling_relation_sigma} and \ref{eq.: Scaling_relation_rcut}). Normalizations are computed at the magnitude of the BCG-N ($mag^{ref}_{\mathrm{F160W}}=17.02$). Red distributions refer to the \qth\ reference lens model; results from the previous model by B19 model are in blue. Colored contours encompass the 1, 2, 3 $\sigma$ confidence levels; the vertical solid and dashed lines correspond to the 50-th, 16-th and 84-th percentiles of the marginalized distributions. The 1 and 2 $\sigma$ black dashed contours refer to the \qth\ model with $r_{core}^{ref}=0.05\arcsec$, instead of $r_{core}^{ref}=1\arcsec \times 10^{-4}$ of the reference model. 
	The green and magenta lines are $\sigma_0\mbox{-}r_{cut}$ curves with constant projected mass, within an aperture of $R= 1\arcsec$ $(=5.34$ kpc at $z=0.396$), for a circular dPIE profile. The mass values from bottom to top are quoted in the legend. Green curves refer to $r_{core}^{ref}=1\arcsec \times 10^{-4}$, magenta curves refer $r_{core}^{ref}=0.05\arcsec$ (as in B19).
	 }
	\label{fig:sigmarcut}
\end{figure}

 In \Fig\ref{fig:sigmarcut}, we show the posterior distribution of the reference central velocity ($\sigma_0^{ref}$) and the reference truncation radius ($r_{cut}^{ref}$) of the cluster member scaling relations, which clearly displays the same degeneracy discussed above. The constraints on these two parameters obtained for the new reference model are compared with the results of B19, which also used a kinematic prior in the lens models. We note that the normalization of the scaling relations ($\sigma_0^{ref}$) in the new model is consistent with B19 at $1\mbox{-}\sigma$ level, whereas we find a difference of $\sim\! 30$ kpc in the normalization of the truncation radius. It is worth noting that such a difference could not be detected without using galaxy kinematics to constrain the scaling relations in the lens model, because in this case the $\sigma_0\mbox{-}r_{cut}$ degeneracy is significantly larger (see B19). 
 
 Since the B19 model assumed a larger value of the reference core radius of the subhalo dPIE profiles ($r_{core}^{ref}=0.05\arcsec$ compared to $1\arcsec\times 10^{-4}$ in \qth), we re-optimize the lens model changing the core radius parameter. The result, illustrated in \Fig\ref{fig:sigmarcut} (dashed confidence contours), shows that $\sigma_0$ scales as expected with the larger core radius, based on \Eq\ref{Eq.:mass_dPIE}, however $r_{cut}$ does not change significantly. Therefore, the larger truncation radii of the subhalos obtained with the new model are not due to inherent degeneracies of the dPIE parametrization with $r_{core}^{ref}$. 

Considering the dependence of the total mass of the subhalos on their values of velocity dispersion and cut radius (\Eq\ref{Eq.:mass_dPIE}), the cluster members in the \qth\ model are characterized by larger masses (by about a factor of two) compared to those in the model of B19. Based on the previous model, \cite{Meneghetti_2020} found that cluster members in \CL\ have a significantly larger probability to produce GGSL compared to galaxy-scale subhalos in $\Lambda$CDM numerical simulations. We have verified that the GGSL probability obtained from the \qth\ model is $\sim30\%$ higher than that measured from the B19 model. Thus, our new results confirm the tension reported by \cite{Meneghetti_2020} between the observations of the inner structure of cluster galaxies and the theoretical expectations in the framework of the $\Lambda$CDM model. 

The constraints on the mass distribution of the subhalo population with lens models can be further improved by taking into account the measured velocity dispersion for each member galaxy, thus including the intrinsic scatter of the  $\sigma\mbox{-}L$ scaling relation \citep{Bergamini_2020}, which is neglected in the current lens models.  We leave to a future work a further analysis of the subhalo component, which fully exploits strong lensing and galaxy kinematics constraints. 

\section{Conclusions}
\label{sec:conclusions}
We have presented a new high-precision strong lens model for the galaxy cluster MACS~J0416.1$-$2403 at $z=0.396$, which takes advantage of the MUSE deep lensed field of 17.1h, carried out in the northeast region of the cluster \citep{Vanzella_2020}. By combining this deep pointing with a careful re-inspection of the HST images, we have identified 82 additional spectroscopic multiple images, compared to the previous catalog published in C17, resulting in a new sample of 182 images from 48 different background sources. In this new sample, we have added several multiply lensed clumps, belonging to resolved extended sources, which are particularly efficient in constraining the position of the critical lines in their vicinity. Moreover, we have extended the sample of cluster members to 213 galaxies, 171 of which are spectroscopically confirmed (27 more than in our previous catalog).

By exploiting new galaxy spectra with high signal-to-noise ratio extracted from the MDLF, we have measured the inner stellar kinematics of 64 cluster members (15 more than in the previous analysis by B19), down to $m_{F160W}\simeq 22$. We have used the newly measured velocity dispersions to constrain the scaling relations of the subhalo population in the lens models as in B19. The cluster-scale mass distribution is modeled with a number of DM-dominated halos in addition to the hot-gas component traced by the Chandra X-ray data. 

Among the four lens models in our study, we have selected a reference model, which best reproduces the positions of the observed multiple images (smallest $\Delta_{rms}$) and the observed $\sigma\mbox{-}mag$ scaling relation for cluster members (see \T\ref{table:lens_models}).

We can summarize the results obtained from our reference \CL\ lens model as follows:

   \begin{enumerate}
      \item Despite the large number of observed multiple images, the new reference lens model yields $\Delta_{rms}=0.40\arcsec$, thus reducing the value obtained by B19 by a third. The model is particularly accurate  in the MDLF region of \CL\ ($\Delta_{rms}^{NE}=0.37\arcsec$, 0.1\arcsec\ smaller than in the southern field), where the newly identified images are located.
       
      \item We have studied the robustness of the magnification maps derived from our reference lens model. We have computed the uncertainties on the absolute magnification ($\left| \mu \right|$) of multiple images on their predicted positions by sampling the $\left| \mu \right|$ distributions from the MCMC chains of the lens model. We have studied how the relative error on the magnification, $\Delta \mu / \left| \mu \right|$, varies with the magnification and with the distance of the multiple images from secondary critical lines associated to cluster galaxies. We find that the relative error remains within $\sim\! 10\%$ at $\left| \mu \right|\lesssim 10$, at distances beyond 2\arcsec from the center of member galaxies, while it increases up to $\sim\! 100\%$ close to critical lines ($\left| \mu \right|\gtrsim 20$). 
      \item We have investigated the ability of the new lens model to reproduce the inner fine structure of extended sources by defining a new metric, which is sensitive to the gradients of the deflection field. We find that the new model can predict the relative distances and orientations of pairs of multiply imaged clumps, inside well resolved sources in the vicinity of critical lines, with errors less than $0.33\arcsec$ and $5.9^{\circ}$, respectively, for 90\% of image pairs. 
      This analysis lends support to the interpretation presented by \citet{Vanzella_2020} of the nature of highly magnified clumps in resolved lensed galaxies in the MLDF. They range from young massive star clusters (e.g., in Sys-14, see also \citealt{Vanzella_ID14}), to barely resolved knots, with sizes and luminosities similar to what observed in local star clusters (e.g., Sys-12). 
      
      \item Using the newly constrained mass distribution of the subhalo component, we confirm the results of \cite{Meneghetti_2020}, who find that the probability of producing GGSL events in several clusters, including \CL, is approximately a factor of ten higher than the theoretical predictions based on $\Lambda$CDM numerical simulations.
      
   \end{enumerate}
   
\noindent The new lens model of \CL\ presented here, together with the new catalog of cluster members and the updated catalog of multiple images presented in \citet{Vanzella_2020}, are made publicly available\footnote{The new and previous lens models are available at \href{http://www.fe.infn.it/astro/lensing/}{www.fe.infn.it/astro/lensing}.}. 
    
\begin{acknowledgements}
      This project is partially funded by PRIN-MIUR 2017WSCC32. PB acknowledges financial support from ASI through the agreement ASI-INAF n. 2018-29-HH.0. MM acknowledges support from the Italian Space Agency (ASI) through contract ``Euclid - Phase D'' and from the grant PRIN-MIUR 2015 ``Cosmology and Fundamental Physics: illuminating the Dark Universe with Euclid''. FC acknowledges support from grant PRIN-MIUR 20173ML3WW\_001. CG acknowledges support by VILLUM FONDEN Young Investigator Programme through grant no. 10123. We acknowledge funding from the INAF ``main-stream'' grants 1.05.01.86.20 and  1.05.01.86.31. GBC acknowledge funding from the European Research Council through the Grant ID 681627-BUILDUP, the Max Planck Society for support through the Max Planck Research Group for S. H. Suyu and the academic support from the German Centre for Cosmological Lensing..
\end{acknowledgements}


\bibliographystyle{aa}
\bibliography{bibliography}

\end{document}